\documentstyle[aps,mytools,epsfig%
]{revtex}

\begin{document}

\title{\Large\bf\sc The Meson Spectrum in a Covariant Quark Model}

\author{Ralf Ricken\thanks{e-mail: {\tt ricken@itkp.uni-bonn.de}}, Matthias
    Koll, Dirk Merten, Bernard Ch. Metsch, Herbert R.
    Petry \\\vspace*{0.25cm}
{\sl Institut f\"ur Theoretische Kernphysik,}\\  
{\sl Nu{\ss}allee 14--16, D--53115 Bonn,\\ 
Germany} }

\date{\today}

\setlength{\parskip}{1ex}
\maketitle

\begin{abstract}
Within the framework of the instantaneous Bethe-Salpeter equation, we present a detailed analysis of light meson spectra with respect to various parameterizations of confinement in Dirac space. Assuming a linearly rising quark-antiquark potential, we investigate two different spinorial forms (Dirac structures), namely $\frac{1}{2}(\Id\otimes\Id - \gamma^0\otimes\gamma^0)$ as well as the $U_A(1)$-invariant combination $\frac{1}{2}(\Id\otimes\Id - \gamma^5\otimes\gamma^5 - \gamma^\mu\otimes\gamma_\mu)$, both providing a good description of the ground state Regge trajectories up to highest observed angular momenta. Whereas the first structure is slightly prefered concerning numerous meson decay properties (see \cite{pap41}), we find the $U_A(1)$-invariant force to be much more appropriate for the description of a multitude of higher mass resonances discovered in the data of the {\sc Crystal Barrel} collaboration during the last few years. Furthermore, this confinement structure
has the remarkable feature to yield a linear dependence of masses on their radial
excitation number. For many experimental resonances such a trajectory-like behaviour
was observed by Anisovich {\it et al.} We can confirm that almost the same slope occurs for
all trajectories. 
Adding the $U_A(1)$-breaking instanton induced 't Hooft interaction we can compute the pseudoscalar mass splittings with both Dirac structures and for the scalar mesons a natural mechanism of flavour mixing is achieved. In the scalar sector, the two models provide completely different ground state and excitation masses, thus leading to different assignments of possible $\bar q q$ states in this region. The scalar meson masses calculated with the structure $\frac{1}{2}(\Id\otimes\Id - \gamma^5\otimes\gamma^5 - \gamma^\mu\otimes\gamma_\mu)$ are in excellent agreement with the K-matrix poles deduced from experiment by Anisovich and coworkers.   
\end{abstract}


\section{Introduction}
In the last few years, evaluations of the {\sc Crystal Barrel} $\bar p
N$-annihilation data set provided a lot of new meson resonances in the mass
region $1000 - 2400$ MeV \cite{pap01,pap02,pap04,pap05,pap06,pap07,pap08}. The experiment was performed with the {\sc Crystal Barrel}
detector at LEAR where antiproton-proton annihilation into $\pi^0\eta\eta$,
$\pi^0\pi^0\eta$, $3 \pi^0$, $4 \pi^0$, $\pi^0\eta$ and $\pi^0\eta\pri$ has
been studied up to an incident beam momentum of $1.94$ GeV/$c$.

From a theoretical point of view, higher excitation
resonances in the meson mass spectrum are of great interest because they reflect
the underlying confinement structure at the quark level. However, from first QCD
principles especially the spinorial Dirac structure of the confinement is largely 
unknown. Therefore a phenomenological study of different Dirac
structures in a fully relativistic framework seems worthwhile. From earlier works
\cite{pap09,pap10} it is known
that a linearly rising confinement potential with a Dirac structure of the form
$\Id\otimes\Id - \gamma^0\otimes\gamma^0$ provides a very good description of
the Regge trajectories up to total angular momentum $J = 6$. However, the suitability of a confinement parameterization should not only be linked to the bulk of experimental ground state masses in each channel but also to their corresponding radial excitations. For that, a multitude of well established higher resonances has to be known. Unfortunately this is not the case so far, although especially during the last two years the situation has improved. For instance, while the {\sc Particle Data Group} 98 (PDG 98 \cite{pap11}) did not list any radial excitations for the isovector states $a_1(11^{++})$, $a_2(12^{++})$, $a_4(14^{++})$ and
$b_1(11^{+-})$, in their latest issue (PDG 00 \cite{pap47}) they state three new such resonances: the vector meson $a_1(1640)$ observed by the authors in \cite{pap02} (a similar resonance, the $a_1(1700)$, is stated in \cite{pap03}), and two tensor mesons, the $a_2(1660)$ \cite{pap05,pap06} and the $a_2(1750)$ \cite{pap48,pap49}. Furthermore a lot of resonances have been found in the last few years, which do not appear in the latest PDG-listing. For example: $a_0(2025)$ \cite{pap01}, $a_1(2100)$ \cite{pap04}, $a_1(2340)$ \cite{pap04}, $a_2(2100)$ \cite{pap04}, $a_4(2260)$ \cite{pap04}, $\eta_2(2040)$ \cite{pap07}, $\eta_2(2300)$ \cite{pap07}, $f_1(1700)$ \cite{pap07},
$f_1(2340)$ \cite{pap07}, $f_4(2320)$ \cite{pap07}. We compare these new
experimental resonance positions with the eigenvalue spectrum
calculated with the Dirac structure $\Id\otimes\Id -
\gamma^0\otimes\gamma^0$ (model ${\cal A}$) on the one hand and the structure $\Id\otimes\Id - \gamma^5\otimes\gamma^5 -
\gamma^\mu\otimes\gamma_\mu$ (model ${\cal B}$) on the other hand. Whereas the first structure
produces excitation masses which overestimate the corresponding experimental
masses up to 400 MeV, the second structure yields a remarkably good
agreement with the newly observed resonances. Here, deviations are in general
less than 100 MeV. Furthermore, in model ${\cal B}$ the squared masses of the resonances show, in contrast to model ${\cal A}$, a linear dependence on their radial excitation number, $M^2 \propto n$, very similar to the behaviour of many experimental resonances recently observed by A. V. Anisovich, V. V. Anisovich and A. V. Sarantsev \cite{pap08}.

It turned out that the comparatively strong coupling of
positive and negative energy components of the Salpeter amplitudes in model ${\cal B}$ is responsible for
this desired lowering of the excited meson masses. We will illustrate this effect by carrying out the
nonrelativistic reduction of the full Salpeter equation for both
models. Whereas even in this limit the above mentioned coupling does not vanish for
model ${\cal B}$, this reduction provides the usual Schr\"odinger equation for model ${\cal A}$,
{\it i.e.} the complete decoupling of positive and negative energy components. Indeed,
model ${\cal A}$ reduces in this limit to a particular version of the nonrelativistic
quark model (NRQM) that provides a satisfactory description of both
meson and baryon ground state masses \cite{pap14}. 
Of course, due to the absence of the negative energy components, the NRQM in \cite{pap14} fails for higher resonances as does the often cited model of Godfrey and Isgur \cite{pap15}.

The relativistic Salpeter framework presented here contains {\it
  a priori} all small desired spin-orbit terms in order to describe the small mass splittings between states that can
be attributed to the same orbital angular momentum multiplets. Examples are
$a_1(1260)$ and $a_2(1320)$, $f_1(1285)$ and $f_2(1270)$ and $K_2(1770)$ and $K^*_3(1780)$. One should emphasize that the low
positions of the $a_0(980)$ and $f_0(980)$ cannot be explained by these
intrinsic spin-orbit terms alone. In order to describe these splittings an additional residual interaction has to be adopted. We will use an instanton induced effective quark interaction discovered by 't Hooft \cite{pap16}, which also accounts for the correct vector-pseudoscalar and $K$-$\pi$-$\eta$-$\eta\pri$ mass splitting. Whereas in the nonrelativistic model
\cite{pap14} this interaction only acts for pseudoscalar mesons, in the fully
relativistic Salpeter model \cite{pap17,pap18} it also acts in the scalar
sector and provides a possible interpretation of the still unknown scalar
ground state nonet \cite{pap19}. However, as we will see this interpretation strongly depends
on details of the confinement force, in particular on whether the used Dirac structure induces additional spin-orbit
forces or not. In fact, as the Dirac structure $\Id\otimes\Id - \gamma^5\otimes\gamma^5 - \gamma^\mu\otimes\gamma_\mu$ generates such additional spin-orbit terms, model ${\cal B}$ provides a completely different result for the scalar mass spectrum compared to the earlier computations in model ${\cal A}$ which produced an almost flavour singlet $f_0$-state at approximately 1 GeV and the flavour octet states $f_0$, $K^*_0$, $a_0$ in a mass region around 1.4 GeV (see \cite{pap19}): In model ${\cal B}$, the calculated scalar ground state masses (and also their radial excitations) are roughly 200--300 MeV lighter. Moreover, they show a remarkable agreement with a lot of K-matrix-poles deduced by V. V. Anisovich and coworkers \cite{pap22,pap23,pap24,pap43,pap44,pap45} from experiment.

This paper is organized as follows: In section II we briefly comment on the Bethe-Salpeter equation for a quark-antiquark bound state and display the approximations (instantaneous approximation, free quark propagators) that lead to the Salpeter equation, which constitutes the basic equation of our model. Section III is devoted to an extensive discussion of the effects of the model interactions adopted on the description of the experimental mass spectrum: In part A, focusing on the confinement force alone, we compare the resulting bound state masses of model ${\cal A}$ and model ${\cal B}$ to the complete $J > 0$ mass spectrum and especially to the recently observed higher resonances quoted above. The nonrelativistic reduction of both models is presented in part B of section III, where we also define the positive and negative energy components of the Salpeter amplitude. A brief discussion concerning spin-orbit effects in both models is given in part C. In the last part of section III, we focus on the pseudoscalar and scalar mesons by adding the residual 't Hooft interaction which only acts for mesons with vanishing total angular momentum. There we will also comment on a set of K-matrix poles found by V. V. Anisovich and coworkers \cite{pap22,pap23,pap24,pap43,pap44,pap45} by comparing to our results. Finally we give a summary and conclusion in section IV.  

\section{A covariant quark model in the instantaneous Bethe-Salpeter approach}
In quantum field theory a quark-antiquark bound state with four-momentum $P$ and mass $M$, $M^2 = P^2$, is described by the Bethe-Salpeter (BS) equation for two fermions \cite{pap12}. In momentum space, this equation reads: 
\begin{eqnarray}
  \label{BS-Gleichung}
  \chi^{P} (p) = \: S_1^F (\frac{P}{2}+p)  
  \int \frac{d^4 p'}{(2\pi)^4} \left[-i K(P, p, p') \chi^{P} (p')\right] S_2^F(-\frac{P}{2}+p), 
\end{eqnarray}
where $p$ is the relative four-momentum between the quark and the antiquark, $K$ denotes the infinite sum of their irreducible interactions and the corresponding full Feynman propagators are labeled by $S^F_1$ and $S^F_2$, respectively. The BS amplitude $\chi^P$ is defined in coordinate space as the time-ordered product of the quark and the antiquark field operator between the bound state $|P\rangle$ and the vacuum:
\begin{eqnarray}        
  \chi ^{P}_{\alpha\beta} (x_1, x_2) &:=&  
  \left\langle\: 0 \: \left |
      T \:\psi_\alpha^1 (x_1)\bar\psi^2_\beta (x_2)\right |\:P
    \:\right\rangle = e^{-iP\cdot(x_1 + x_2)/2} \int\frac{d^4 p}{(2\pi)^4}
  e^{-ip\cdot(x_1 - x_2)} \chi ^{P}_{\alpha\beta} (p) ,
\end{eqnarray}
where $\alpha$ and $\beta$ are multi-indices for the Dirac, flavour and colour degrees of freedom.

Since in general the interaction kernel $K$ and the full quark propagators
$S^F$ are unknown quantities we make two (formally covariant) approximations:
\begin{itemize}
\item The propagators are assumed to be of the free form $S^F_i = i
  (\;\not\!\! p - m_i + i\epsilon\;)^{-1}$ with effective constituent quark masses
  $m_i$ that we use as free parameters in our model.
\item It is assumed that the interaction kernel only depends on the components
  of $p$ and $p\pri$ perpendicular to $P$, {\it i.e.} $K(P,p,p\pri) = V(p_{\perp P},
  p\pri_{\perp P})$ with $p_{\perp P} := p - (p\cdot P/P^2)P$ (instantaneous
  approximation).
\end{itemize}
Integrating in the bound state rest frame over the time component
$p^0$ and
introducing the equal-time (or Salpeter) amplitude
\begin{eqnarray}
  \Phi (\vec p) := \int\frac{dp^0}{2\pi} \:\left.\chi^P (p^0, \vec p)
  \right|_{P=(M, \vec 0)} 
  = \int\frac{dp_{\parallel P}}{2\pi} \:\left.\chi ^P
    (p_{\parallel P}, p_{\perp P})\right|_{P=(M, \vec 0)}\; , 
\end{eqnarray}
we end up with the Salpeter equation~\cite{pap13}, which constitutes
the basic equation of our model:
\begin{eqnarray}\label{eq:SE}
  \Phi (\vec p) &=& + \:\Lambda_1^-(\vec p) \gamma ^0 \left[ 
    \int\frac{d^3 p'}{(2\pi)^3} \frac{V(\vec p, \vec p\, ') 
      \Phi (\vec p\, ')}{M+\omega _1 + \omega _2} \right] 
  \gamma ^0\Lambda _2^+ (-\vec p)\nonumber \\
  & & -\: \Lambda_1^+(\vec p) \gamma ^0 \left[ \int\frac{d^3
      p'}{(2\pi)^3}\frac{ V(\vec p, \vec p\, ') \Phi (\vec p\, ')}{M-\omega _1 -
      \omega _2} \right] \gamma ^0\Lambda _2^-  (-\vec p)\; .
\end{eqnarray}
Here $\Lambda _i^\pm(\vec p) = (\omega_i \pm \gamma^0(\vec \gamma 
\vec p + m_i))/2\omega_i$ are projectors on positive and negative energy solutions of the Dirac equation
and 
$\omega_i=\sqrt{\vec p^2 + m_i^2}$ denotes the kinetic energy of the quarks.
  
The simultaneous calculation of the meson masses $M$ and the
Salpeter amplitudes $\Phi$ results by solving the corresponding
eigenvalue problem of eq. (\ref{eq:SE}) with an adequate
potential ansatz (see \cite{pap18} for details).
\section{Model Interactions and Meson Mass Spectra}
The global structure of the experimental mass spectrum reflects a linearly rising confinement force which not only produces the Regge trajectories $M^2 \propto J$ but also the energy mass gaps between the radial excitation states. Furthermore, the confinement force should be flavour symmetric because one finds for every isovector state an energetically degenerate isoscalar partner in the experimental mass spectrum. The best known example is of course $\rho(770)$ and $\omega(782)$, but also $h_1(1170)$ and $b_1(1235)$ and many other pairs, up to $a_6(2450)$ and $f_6(2510)$. The pseudoscalar mesons $\pi$, $\eta$, $\eta\pri$ exhibit a mass splitting which is of course not compatible with this rule; therefore one has to introduce a flavour dependent residual interaction in order to get an appropriate description for these mesons. We will use the instanton induced effective quark interaction discovered by 't Hooft \cite{pap16} which in the present formulation acts for mesons with vanishing total angular momentum ($J = 0$) only. In part A of this section, we will focus on the confinement force alone and compare the resulting bound state masses of model ${\cal A}$ and model ${\cal B}$ to the corresponding experimental resonances with $J > 0$. Presenting the nonrelativistic reduction of both models in part B and discussing spin-orbit effects in part C, the residual 't Hooft interaction will be added in part D in order to cover the pseudoscalar and scalar mesons.
\subsection{Confinement Potential and Mesons with $J > 0$}
In this subsection, we will discuss two different versions for the confinement force. These two models differ by their Dirac structures $\Gamma\otimes\Gamma$ whereas the coordinate space dependence is chosen to be linear in both parameterizations:
\begin{eqnarray}
\int\frac{d^3 p\pri}{(2\pi)^3} V_C(\vec p, \vec p\:') \Phi(\vec p\:') = \int\frac{d^3 p\pri}{(2\pi)^3}{\cal V}_C((\vec p - \vec p\:')^2) \Gamma\Phi(\vec p\:')\Gamma \; .
\end{eqnarray}
Here ${\cal V}_C((\vec p - \vec p\:')^2)$ is the Fourier transform of the linearly rising potential ${\cal V}(|\vec x_q - \vec x_{\bar q}|) = a_c + b_c \cdot |\vec x_q - \vec x_{\bar q}|$; the confinement offset $a_c$ and its slope $b_c$ are free parameters of our model. So, for given Dirac structure and constituent quark masses in a physically reasonable range ($m_n \approx 300 - 400$ MeV, $m_s \approx 500 - 600$ MeV) the calculation of the complete $J > 0$ mass spectrum only depends on these two parameters. In order to fix them, we perform a fit to the experimental Regge trajectories and, after some fine tuning of the quark masses, we end up with the parameters of model ${\cal A}$ and model ${\cal B}$ shown in table \ref{tab:Parameters}.\footnote{The parameters of the 't Hooft interaction are fixed in the pseudoscalar sector and will be discussed in part D of this section.}
\subsubsection{Dirac Structure $\Gamma\otimes\Gamma = \frac{1}{2}(\Id\otimes\Id - \gamma^0\otimes\gamma^0)$ (Model ${\cal A}$)}
\begin{figure}[h]
\input{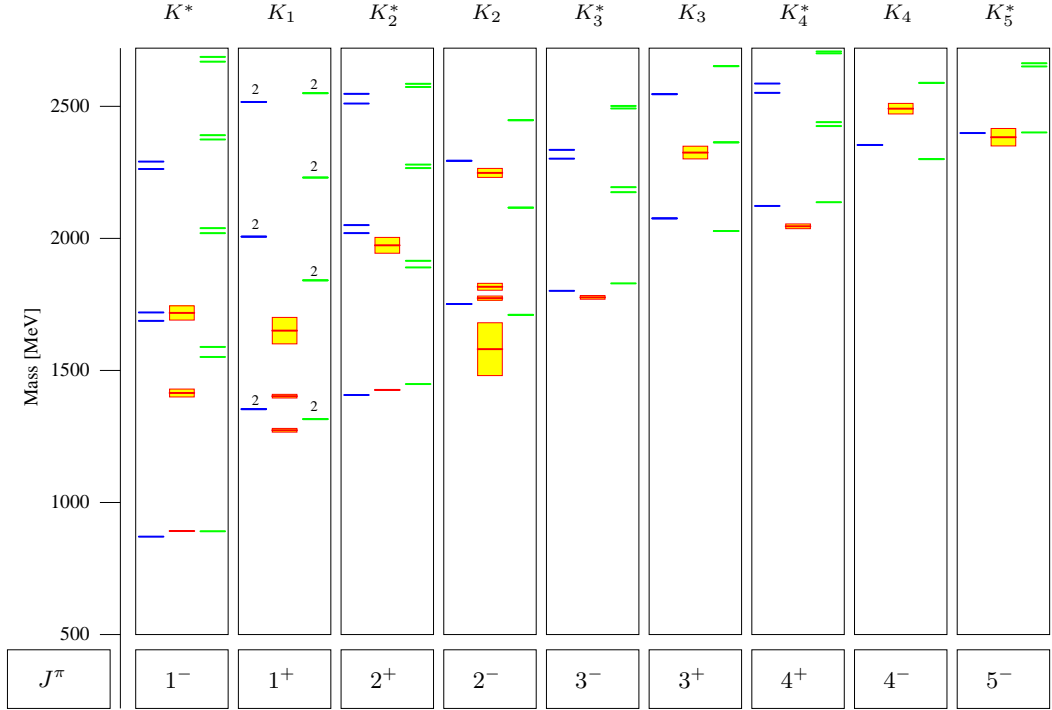}
\centering
\newline
\caption{Strange meson spectrum ($J > 0$). In the middle of each column the experimental resonance positions [47] and their errors are indicated by lines and shaded areas; the lines in the left and right part of each column represent the calculated masses in model ${\cal A}$ and in model ${\cal B}$, respectively. Note that the calculated $K_1$ states are each 2-fold degenerate for spin $S = 0$ and $S = 1$, indicated by '2'.} 
\label{fig:spect05}
\end{figure}

This combination of a scalar and a timelike vector Dirac structure has already been used in earlier works \cite{pap09,pap10,pap19} where a very good description of the Regge trajectories $M^2 \propto J$ was achieved. The complete ($J > 0$) spectrum up to $J = 6$ with all its radial excitations up to 2.5 GeV is shown in figs. (\ref{fig:spect0})--(\ref{fig:spect05}) where none of the new experimental resonances mentioned in the introduction were used in the fit but only the masses listed by the {\sc Particle Data Group} \cite{pap47}. Consequently a quantitative statement concerning the quality of the higher excitation calculations is not possible on the basis of these data alone. Especially, the higher radial excitations of the isovector states $b_1$, $a_1$ or of their isoscalar partners $h_1$, $f_1$ are not contained in \cite{pap47}. Fortunately, the situation has changed illustrated by the data shown in table \ref{tab1} and table \ref{tab2}. Here we have listed several new resonances which all have been found during the last few years by various groups and collaborations \cite{pap01,pap02,pap03,pap04,pap05,pap06,pap07}. 

A comparison of the newly observed resonances with our calculations shows that the Dirac structure $\Id\otimes\Id - \gamma^0\otimes\gamma^0$ (model ${\cal A}$) produces masses that are roughly 150 - 350 MeV too high compared to the experimental values. The relativized quark model of Godfrey and Isgur \cite{pap15} provides a similar tendency although their deviations are smaller. 
However, this does not exclude a quarkonium interpretation of these states around 1700 MeV and 2100 MeV. Starting with the $a_1(1640)$ \cite{pap02} or $a_1(1700)$ \cite{pap03} as the well established 2P $n\bar n$ state, a $\bar q q$ classification is much more natural than other conceivable interpretations of these resonances: Then the mass centroid of the 2P multiplets is around 1700 MeV and the multiplet partners of the $a_1$ are expected nearby in mass provided that splittings due to spin-orbit and tensor forces are small even for orbital angular momentum $L = 1$ (P-wave) and $L = 3$ (F-wave). Indeed, the {\sc Crystal Barrel} collaboration observed a new state, the $a_2(1660)$, in the reaction $\bar p p$ $\rightarrow$ $\pi^0 \eta\eta$ at 1.94 GeV/$c$ \cite{pap06}. A similar resonance has been reported in \cite{pap05} where the authors performed a combined K-matrix analysis of the GAMS, {\sc Crystal Barrel} and BNL data. In addition there is also experimental evidence for the next higher multiplet: The authors in \cite{pap04} studied the process $\bar p p \rightarrow f_2(1270)\pi$ in the mass range 1960-2410 MeV and they observed a 1$^{++}$ and a 2$^{++}$ resonance around 2100 MeV which can be identified with the $a_1(2100)$ and the $a_2(2100)$, respectively. They also found evidence for a $1^{++}$ resonance at 2340 MeV, ($a_1(2340)$), a $4^{++}$ resonance at 2260 MeV, ($a_4(2260)$), and two 3$^{++}$ resonances at 2070 MeV and 2310 MeV, ($a_3(2070)$ and $a_3(2310)$). However, as already mentioned, neither model ${\cal A}$ nor the model of Godfrey and Isgur produces masses around 1700 MeV and/or 2100 MeV which fit to the above quantum numbers. In more detail, model ${\cal A}$ predicts the first radial excitation of the (pure P-wave) $a_1$ at 1876 MeV and the second radial excitation at 2374 MeV. For the first radial excitation of the $a_2$, the model provides the dominantly F-wave state at 1879 MeV {\it below} the dominantly P-wave state at 1931 MeV. 

In the isoscalar sector (see table \ref{tab2}), the classification of calculated and experimental resonances is more difficult due to the doubling of states by the additional $\bar s s$ pair. However, the approximate flavour symmetry of the inter-quark forces suggests to expect the first nonstrange radial excitation masses of the $f_1$ and the $f_2$ also around 1700 MeV and the corresponding second radial excitations around 2100 MeV. The third nonstrange radial excitation of the $f_1$ should appear in a mass range of about 2300-2400 MeV and the first nonstrange radial excitation of the $f_4$ around 2300 MeV. Moreover the isoscalar ground state partner of the $a_3$ should appear around 2000 MeV and its first radial excitation partner around 2300 MeV. In fact, the authors in \cite{pap07} found evidence for resonances with these quantum numbers. They studied the process $\bar p p \rightarrow \pi^0 \pi^0 \eta$ for beam momenta of 600 - 1940 MeV/$c$, corresponding to center of mass energies 1960-2410 MeV, and found a 1$^{++}$ resonance at 2340 MeV, the $f_1(2340)$ and a 4$^{++}$ resonance at 2320 MeV, the $f_4(2320)$. The masses of the isoscalar partners of the $a_3$ were stated at 2000 MeV, ($f_3(2000)$), and at 2280 MeV, ($f_3(2280)$). As in the isovector sector, we fix the mass centroid of the 2P $n\bar n$ multiplet around 1700 MeV assuming that the $f_1(1700)$ is the isoscalar partner of the $a_1(1700)$. Although the $f_1(1700)$ was below the accessible mass range in \cite{pap07}, the assumption above is supported by the mass degeneracy of the observed $f_1(2340)$ with its isovector partner $a_1(2340)$. Now, due to the flavour independence of the confinement interaction used in our model, the calculated isoscalars are ideally mixed and the nonstrange part of the spectrum is degenerate in mass with the corresponding isovector spectrum. Therefore, also the isoscalar masses calculated in model ${\cal A}$ are too high compared to the experimental values.\\ 
In order to fix the centroid of the 2P $s\bar s$ multiplet, we consider the following well established splittings of $\bar n n$ and $\bar s s$ isoscalar ground states listed by the {\sc Particle Data Group} \cite{pap47}: [$\omega(782)$, $\Phi(1020)$], [$h_1(1170)$, $h_1(1380)$], [$\eta_2(1645)$, $\eta_2(1870)$], [$\omega_3(1670)$, $\Phi_3(1850)$]. These mass splittings of approximately 200-300 MeV leads to an expected centroid of the 2P $\bar s s$ multiplet at roughly 1950-2050 MeV provided that the 2P $n\bar n$ multiplet appears around 1700 MeV as discussed above. Furthermore the {\sc Particle Data Group} \cite{pap47} states two isoscalar 2$^{++}$ resonances, the $f_2(1950)$ and $f_2(2010)$, which fit to this energy region such that an arrangement of one of these states into the 2P $s\bar s$ multiplet seems to be natural. However, the mass calculations in model ${\cal A}$ for the strange part of the isoscalar 2$^{++}$ sector provide a dominantly F-wave $f_2$ at 2148 MeV and a dominantly P-wave $f_2$ at 2165 MeV as can be seen in fig. (\ref{fig:spect0}). Moreover, the model produces the first strange radial excitation state of the $f_1$ at 2129 MeV. So, the strange part of the isoscalar excitation spectrum comes out too high in model ${\cal A}$ just like the nonstrange excitation spectrum discussed above.

In view of these facts, a quarkonium interpretation for the higher mass resonances seems still plausible but obviously can only be achieved with a new confinement force. 
\subsubsection{Dirac Structure $\Gamma\otimes\Gamma = \frac{1}{2}(\Id\otimes\Id - \gamma^5\otimes\gamma^5 - \gamma^\mu\otimes\gamma_\mu)$ (Model ${\cal B}$)}
\begin{figure}[b]
\input{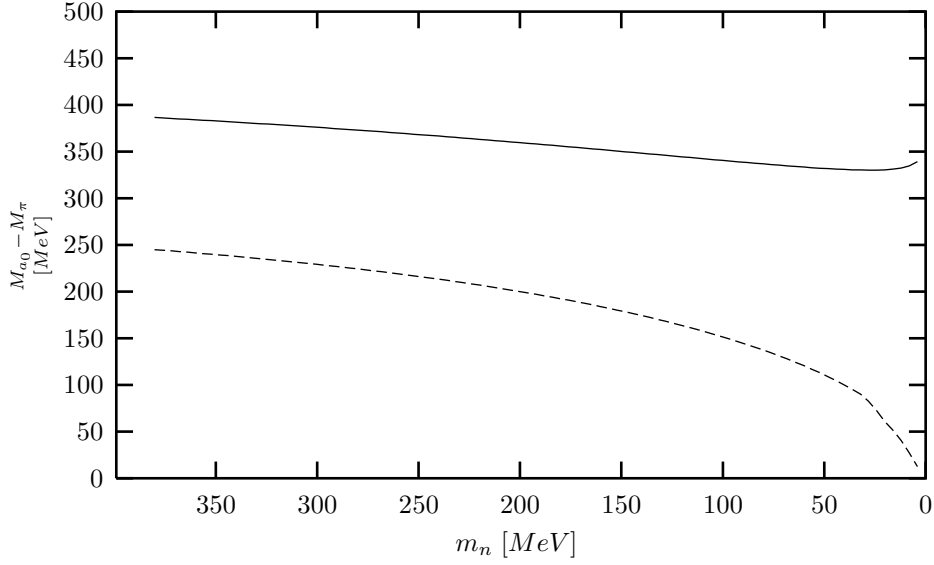}
\centering
\newline
\caption{The $a_0$-$\pi$ ground state mass difference as a function of the constituent quark mass $m_n$ calculated in model ${\cal A}$ (solid line) and model ${\cal B}$ (dashed line). For vanishing constituent quark masses, the $a_0$-$\pi$ mass degeneracy in model ${\cal B}$ is a direct consequence of the $U_A(1)$-invariance of the Dirac structure $\frac{1}{2}(\Id\otimes\Id - \gamma^5\otimes\gamma^5 - \gamma^\mu\otimes\gamma_\mu)$. In model ${\cal A}$ this degeneracy is not observed due to the $U_A(1)$-violating scalar part of the structure $\frac{1}{2}(\Id\otimes\Id - \gamma^0\otimes\gamma^0)$. The calculation was done without the explicit $U_A(1)$-breaking 't Hooft interaction.}\label{fig:m0klsp4diff} 
\end{figure}

From first principles QCD hardly gives any clue on the Dirac structure of confinement. Consequently a phenomenological study of different Dirac structures is needed. The force with $\Id\otimes\Id - \gamma^0\otimes\gamma^0$ has been found by purely phenomenological arguments. Here the aim was to reproduce the global structure of the meson mass spectrum, especially the well established Regge trajectories. Now we focus on some symmetry properties of the strong interaction. For example the spontaneous breakdown of the approximate chiral symmetry of light flavour QCD leads to an effective theory, the well known chiral perturbation theory, which interprets the eight lightest pseudoscalar mesons as Goldstone bosons and fixes their interactions by symmetry requirements (apart from phenomenological coupling constants). Furthermore it is believed that, due to the absence of a ninth light pseudoscalar meson, the axial $U(1)$-symmetry is explicitly broken by instantons. As we implement this by using the instanton induced 't Hooft interaction (see part D of this section) it is interesting to assume a confinement force with axial $U(1)$-{\it invariance} in the present framework. We choose the structure 
\begin{eqnarray}
\Gamma\;\otimes\;\Gamma = \frac{1}{2}(\Id\otimes\Id - \gamma^5\otimes\gamma^5 - \gamma^\mu\otimes\gamma_\mu)\;.
\end{eqnarray}
Due to the $U_A(1)$-invariance, this structure obviously leads to parity doublets in the calculated meson spectrum if one neglects the quark mass terms in the Salpeter equation (see fig. (\ref{fig:m0klsp4diff})).\\ 
The above Fierz invariant combination of a scalar, pseudoscalar and vector part was also investigated by B\"ohm {\it et al.} (see \cite{pap55}) as well as by Gross and Milana (see \cite{pap56}) but, unlike the investigations presented here, without any connection to the complete experimental meson spectrum.

As figs. (\ref{fig:spect0})--(\ref{fig:spect05}) show, this Dirac structure yields ground state Regge trajectories $M^2 \propto J$ that are as good as the corresponding results achieved with the scalar timelike vector force discussed above. However, both models differ essentially in their results for the higher mass resonances: In contrast to model ${\cal A}$, the calculated higher excitation masses in model ${\cal B}$ agree remarkably well with the newly observed resonances (see tables \ref{tab1} and \ref{tab2}). Here, deviations are in general less than 100 MeV. For instance, the first radial excitation of the $a_1$ appears at 1718 MeV (see table \ref{tab1}), which means a lowering of about 160 MeV compared to the corresponding mass calculated in model ${\cal A}$ (1876 MeV). The second radial $a_1$-excitation appears at 2099 MeV, so it is even 275 MeV lighter than the corresponding mass in model ${\cal A}$ (2374 MeV), thus showing excellent agreement with the observed resonance $a_1(2100)$ \cite{pap04}. For the third radial $a_1$-excitation model ${\cal B}$ provides 2412 MeV compared to 2791 MeV in model ${\cal A}$ and 2340 $\pm$ 40 MeV the corresponding value from experiment \cite{pap04}. As can be seen in table \ref{tab1}, not only the $a_1$-excitation masses are lowered in model ${\cal B}$ but also all other higher isovector resonances listed there. Due to the flavour independence of the confinement force, the same effect occurs in the complete nonstrange and strange isoscalar spectrum (see table \ref{tab2}).

\begin{figure}
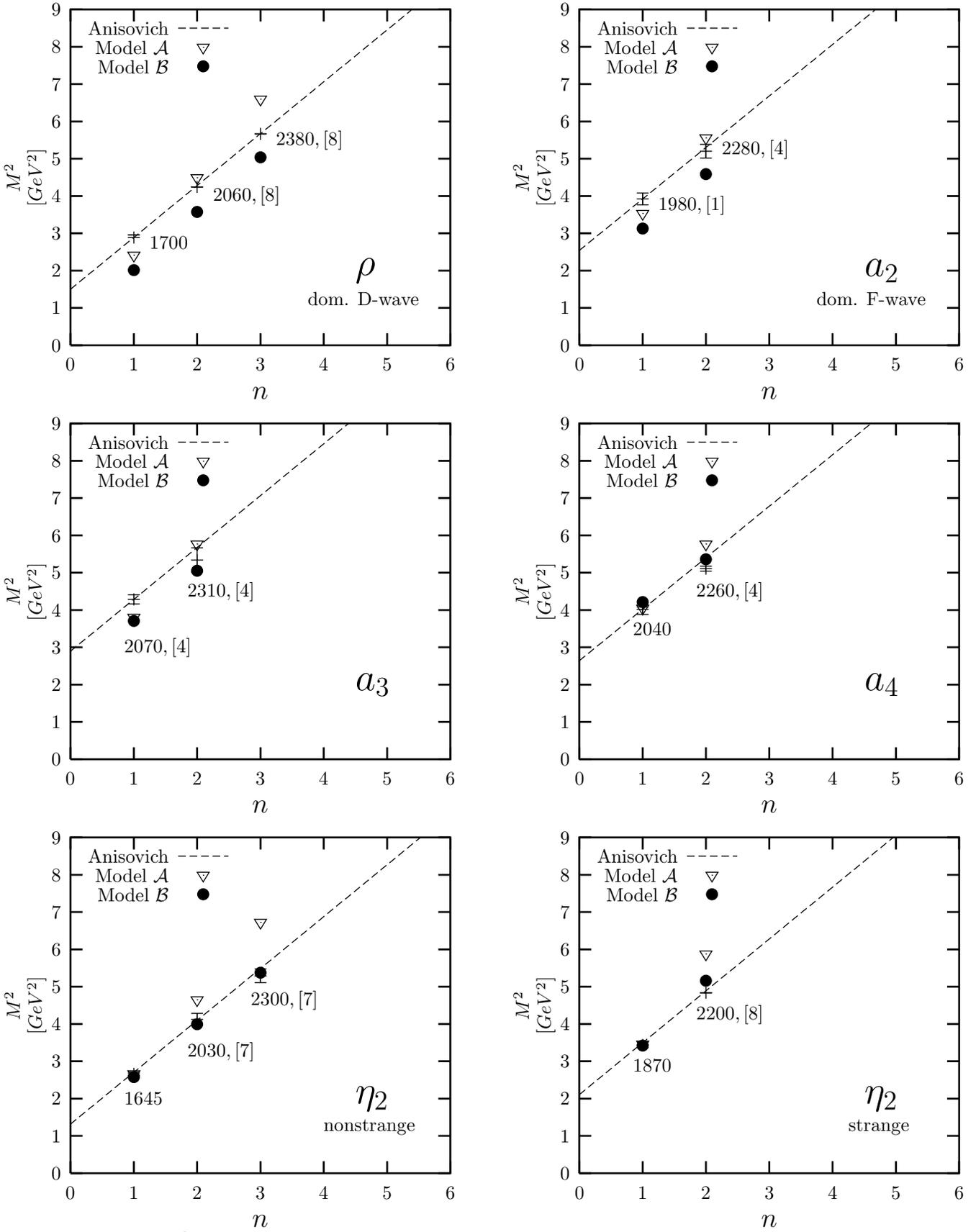

\begin{minipage}[]{17.9cm}

\begin{minipage}[t]{8.5cm}
     \input{MrhoDdomvsn_new.latex}
\end{minipage}
\hspace{0.7cm}
\begin{minipage}[t]{8.5cm}
     \input{Ma2Fdomvsn_new.latex}
\end{minipage}

\begin{minipage}[]{8.5cm}
     \input{Ma3vsn_new.latex}
\end{minipage}
\hspace{0.7cm}
\begin{minipage}[]{8.5cm}
     \input{Ma4vsn_new.latex}
\end{minipage}

\begin{minipage}[b]{8.5cm}
     \input{Meta21vsn_new.latex}
\end{minipage}
\hspace{0.7cm}
\begin{minipage}[b]{8.5cm}
     \input{Meta22vsn_new.latex}
\end{minipage}

\caption{The ($n,M^2$)-trajectories for the states $\rho(11^{--})$,
  $a_2(12^{++})$, $a_3(13^{++})$, $a_4(14^{++})$ and $\eta_2(02^{-+})$. The resonances indicated with the reference number [8] are predictions according to the formula $M^2 = M^2_0 + (n - 1) \mu^2$. The trajectory slopes $\mu^2$ are taken from [8] and are approximately the same for all
  trajectories. The resonances without a reference number are listed in [47]. All other resonances have been indicated with the reference number where they have been seen or predicted.}\label{fig:Mvsn3}
\end{minipage}
\end{figure}

In connection with this general lowering of excitation masses in model ${\cal B}$, we can confirm a recent observation found by A. V. Anisovich, V. V. Anisovich and A. V. Sarantsev \cite{pap08} concerning the systematics of $\bar q q$ states with respect to their radial excitation number. The authors found that for given quantum numbers many mesons fit to linear trajectories as
\begin{eqnarray}\label{eq:M2vsn}
M^2 = M^2_0 + (n - 1)\mu^2
\end{eqnarray}
with $M_0$ the mass of the ground state meson ($n = 1$), $n = 2,3,...$ numerating the radial excitations and $\mu^2$ the trajectory slope parameter whose value is suggested to be in the region $\mu^2 = 1.25 \pm 0.15$ GeV$^2$. Figs. (\ref{fig:Mvsn1}) and (\ref{fig:Mvsn3}) show several resonance positions taken from \cite{pap08} and our corresponding calculations of model ${\cal A}$ and model ${\cal B}$. As expected from the discussion of table \ref{tab1} and table \ref{tab2}, the trajectory slope $\mu^2$ as given in \cite{pap08} is much too flat to parameterize the calculations of model ${\cal A}$ whereas the model ${\cal B}$ calculations fit for many mesons to the formula in eq. (\ref{eq:M2vsn}) with slope $\mu^2 = 1.25 \pm 0.15$ GeV$^2$ remarkably well.\footnote{However, a more detailed study of our calculations with respect to eq. (\ref{eq:M2vsn}) would prefer especially for the $\pi(10^{-+})$- and $a_0(10^{++})$-trajectories (fig. (\ref{fig:Mvsn1})) a somewhat larger slope, namely $\mu^2(\pi(10^{-+})) = 1.59$ GeV$^2$ and $\mu^2(a_0(10^{++})) = 1.56$ GeV$^2$. Then the experimental resonances $\pi(1300 \pm 100)$ and $\pi(1800 \pm 40)$ and the $a_0(2000^{+50}_{-100})$ would fit much better to their corresponding trajectories. Consequently, due to the higher slopes, the parameterization in eq. (\ref{eq:M2vsn}) would lead to higher mass predictions, namely $\pi(2189)$, $\pi(2526)$ and $a_0(1588)$, $a_0(2375)$ instead of $\pi(2070)$, $\pi(2380)$ and  $a_0(1520)$, $a_0(2260)$ as stated in \cite{pap08}. The higher masses then would correspond to the model ${\cal B}$ calculations $\pi(2195)$, $\pi(2496)$ and $a_0(1665)$, $a_0(2395)$. \\
A universal trajectory fit to our mass calculations provides an average slope of $\mu^2 = 1.42 \pm 0.27$ GeV$^2$ which is roughly 14$\%$ larger than the value $\mu^2 = 1.25 \pm 0.15$ GeV$^2$ suggested in \cite{pap08}. However, this deviation should not be overinterpreted because there is no obvious reason to demand an approximately unique slope for all trajectories.}

In summary, the calculated excitation masses of model ${\cal B}$ are in general roughly 150-350 MeV lighter than the corresponding masses of model ${\cal A}$, hence in a much better agreement with the newly discovered experimental resonances. As we will show below, this significant mass lowering is related to the strong coupling between positive and negative energy components of the Salpeter amplitude in model ${\cal B}$.
\subsection{Nonrelativistic Reduction of the Salpeter Equation}
\begin{figure}[b]
\input{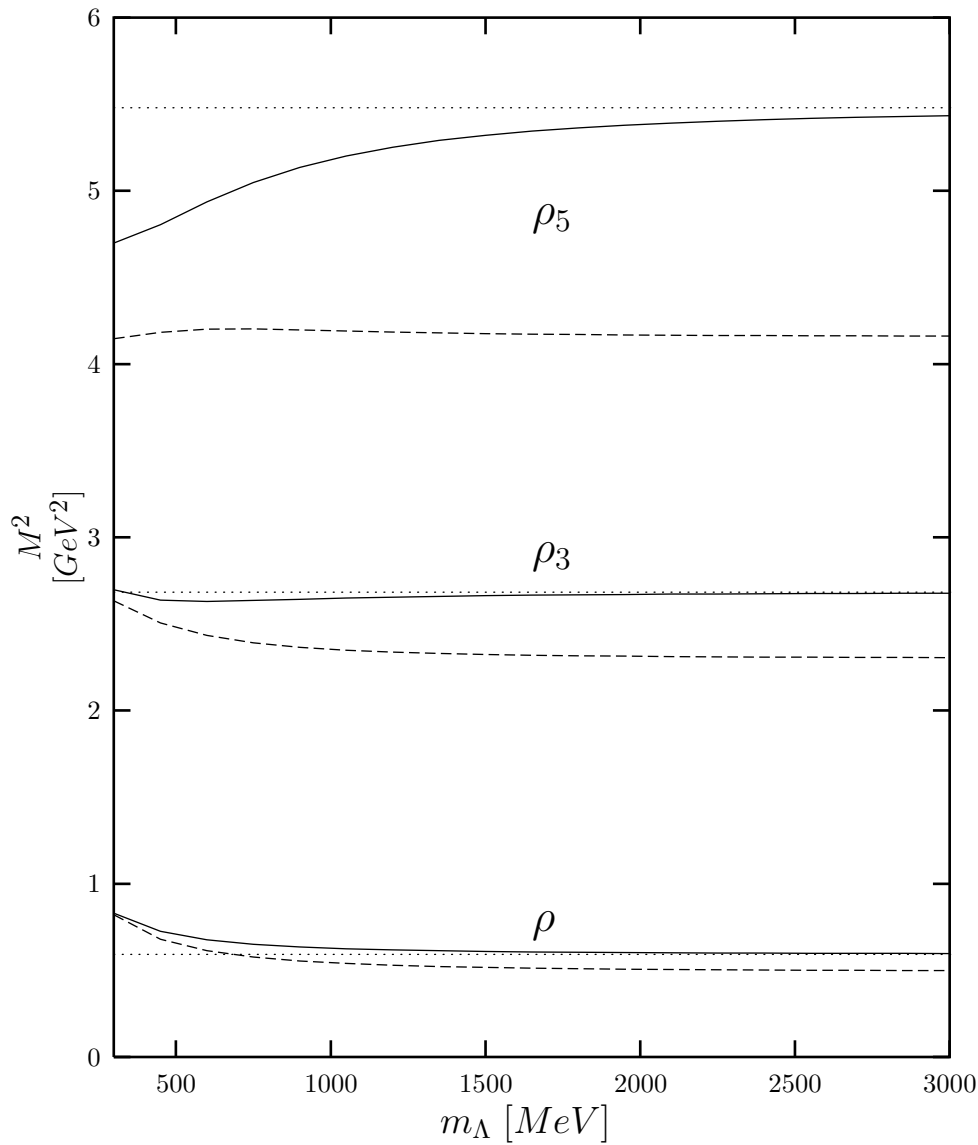}
\centering
\newline
\caption{The squared masses of the isovector Regge trajectory members $\rho$, $\rho_3$ and $\rho_5$ as a function of the constituent quark mass $m_\Lambda$ in the energy projectors (see eq. (5)) calculated in the nonrelativistic (Schr\"odinger equation) model [14] (dotted lines), in the full Salpeter model with Dirac structure $\frac{1}{2}(\Id\otimes\Id - \gamma^0\otimes\gamma^0)$ (solid lines) and with Dirac structure $\frac{1}{2}(\Id\otimes\Id - \gamma^5\otimes\gamma^5 - \gamma^\mu\otimes\gamma_\mu)$ (dashed lines). The calculations were done with the {\it single} parameter set of the nonrelativistic (Schr\"odinger equation) model [14]. Whereas the nonrelativistic reduction ($m_\Lambda \rightarrow \infty$) of the structure $\frac{1}{2}(\Id\otimes\Id - \gamma^0\otimes\gamma^0)$ coincides with this model (see eq. (12)), the negative energy components $\Phi^{--}$ generated by the structure $\frac{1}{2}(\Id\otimes\Id - \gamma^5\otimes\gamma^5 - \gamma^\mu\otimes\gamma_\mu)$ do not vanish in this limit (see eq. (26) $\&$ eq. (27)). They lead to significantly negative mass shifts with respect to the nonrelativistic model calculations. The more the meson mass $M$ differs from the weakly bound state condition $M \approx m_1 + m_2$ (see eq. (8)) the larger are these effects.}\label{fig:rhorho3rho5}
\end{figure}

In this subsection, we will show that the negative energy Salpeter components play an essential role for the description of higher excitation states in the meson mass spectrum. It is known from earlier works (see {\it e.g.} \cite{pap09,pap10}) that these components are very important in calculations of electroweak observables, especially for deeply bound states. There it has been shown that a fully relativistic treatment in the Salpeter framework improves the description of many observables (for example: pseudoscalar decay constants, two photon decay widths) by orders of magnitude. In order to define the negative energy components of the Salpeter amplitude $\Phi$, we look at the nonrelativistic reduction of the Salpeter equation. This reduction of eq. (\ref{eq:SE}) is reached by neglecting all momentum dependent terms in the energy projectors (formally by performing here the limit $m_i\rightarrow\infty$) and by expanding the kinetic energies $\omega_i$ with respect to $|\vec p_i| / m_i \ll 1$:
\begin{eqnarray}
\lim_{m_i\to\infty}\Lambda^{\pm}_i(\vec p) &=& \lim_{m_i\to\infty}\frac{\omega_i \pm \gamma^0 (\vec \gamma \vec p + m_i)}{2 \omega_i} = \frac{1}{2}(\Id \pm \gamma^0) =: {\cal P}^{\pm} \\
\omega_i &=& \sqrt{|\vec p|^2 + m^2_i} \approx m_i + \frac{1}{2}\frac{|\vec p|^2}{m_i} =: \;\tilde \omega_i \mbox{ for } \frac{|\vec p|}{m_i} \ll 1\; .
\end{eqnarray}
Inserting these approximations, the Salpeter equation (4) becomes:
\begin{eqnarray}
\label{eq:redS}
  \Phi (\vec p) &=& + \:{\cal P}^{-}\gamma ^0 \left[ 
    \int\frac{d^3 p'}{(2\pi)^3} \frac{V(\vec p, \vec p\, ') 
      \Phi (\vec p\, ')}{M+ \tilde\omega _1 + \tilde\omega _2} \right] 
  \gamma ^0{\cal P}^{+}\nonumber \\
  & & -\: {\cal P}^{+}\gamma ^0 \left[ \int\frac{d^3
      p'}{(2\pi)^3}\frac{ V(\vec p, \vec p\, ') \Phi (\vec p\, ')}{M- \tilde\omega _1 -
      \tilde\omega _2} \right] \gamma ^0{\cal P}^{-}\; .
\end{eqnarray}
Now, if we write the 4$\times$4-matrices $\Phi$ and $V\Phi$ in block matrix form as 
\[ \Phi(\vec p) =: \left( \begin{array}{lr}\Phi_{11}(\vec p) & \Phi_{12}(\vec p) \\ \Phi_{21}(\vec p) & \Phi_{22}(\vec p)\end{array}\right) \]
and 
\[ V(\vec p,\vec p\:')\Phi(\vec p\:') =: \left( \begin{array}{lr}(V\Phi)_{11}(\vec p, \vec p\:') & (V\Phi)_{12}(\vec p, \vec p\:') \\ (V\Phi)_{21}(\vec p, \vec p\:') & (V\Phi)_{22}(\vec p, \vec p\:')\end{array}\right), \]
where each component $\Phi_{ij}$ and $(V\Phi)_{ij}$, $i,j \in \{1,2\}$ is a 2$\times$2 matrix and apply the projectors ${\cal P}^{\pm}$ in Dirac representation\footnote{Dirac representation of the $\gamma$ matrices: \[ \gamma^0 = \left( \begin{array}{lr} \Id & 0 \\ 0 & -\Id \end{array}\right),\;\; \vec\gamma = \left( \begin{array}{cc} 0 & \vec\sigma \\ -\vec\sigma & 0 \end{array}\right),\;\; \gamma^5 = \left( \begin{array}{lr} 0 & \Id \\ \Id & 0 \end{array}\right)\;. \]} on both sides of eq. (\ref{eq:redS}), we get:
\label{eq:redS2}
\[\left( \begin{array}{cc}0 & \Phi_{12}(\vec p) \\ \Phi_{21}(\vec p) & 0\end{array}\right) = \left( \begin{array}{cc}0 & \left[\int\frac{d^3 p'}{(2\pi)^3} \frac{(V\Phi)_{12}(\vec p, \vec p\, ')}{M - \tilde\omega _1 - \tilde\omega _2}\right] \\ -\left[\int\frac{d^3
      p'}{(2\pi)^3}\frac{ (V\Phi)_{21}(\vec p, \vec p\, ')}{M + \tilde\omega _1 +
      \tilde\omega _2}\right] & 0\end{array}\right)\;. \]
For weakly bound states with $M\approx m_1 + m_2$ one has 
\begin{equation}
\frac{1}{M + \tilde \omega_1 + \tilde \omega_2} \ll \frac{1}{M - \tilde \omega_1 - \tilde \omega_2}
\end{equation}
so that the component $\Phi_{21}$ can be dropped. The above matrix equation then decouples with respect to $\Phi_{12}$ and $\Phi_{21}$ due to the smallness of $\Phi_{21}$. Therefore one can interpret $\Phi^{++} := \Phi_{12}$ as the upper (or positive energy) component of the Salpeter amplitude $\Phi$. Consequently, $\Phi^{--} := \Phi_{21}$ can be interpreted as the lower (or negative energy) component of $\Phi$.\footnote{The special projector structure of the Salpeter equation (4) allows to express the diagonal components $\Phi^{+-} := \Phi_{11}$ and $\Phi^{-+} := \Phi_{22}$ in terms of $\Phi^{++}$ and $\Phi^{--}$ as
\begin{eqnarray}
\Phi^{+-} &=& + c_1 \; \Phi^{++}s - c_2\; s\; \Phi^{--} \\
\Phi^{-+} &=& - c_1 \; \Phi^{--}s + c_2\; s\; \Phi^{++}\; ,
\end{eqnarray}
with the shorthand notations $s = \vec \sigma \vec p$, $c_i$ = $\frac{\omega_i}{(\omega_1 m_2 + \omega_2 m_1)}$ (see \cite{pap18} for details).} One should emphasize that the above approximate decoupling for weakly bound state solutions is a consequence of the Salpeter equation itself and independently from the special form of the potential operator $V$ in Dirac space\footnote{This statement is at least valid for the Dirac structures discussed in the present paper.}. With usual constituent quark masses of 300--400 MeV, one therefore can expect to find different Dirac structures which all produce for example the $\rho$-meson mass around 700--800 MeV with the {\it same} potential parameters. However, for deeply bound states ($M \ll m_1 + m_2$) on the one hand and higher excitation states ($M \gg m_1 + m_2$) on the other hand, neglecting the negative energy components $\Phi^{--}$ is no longer justified. So, in that case even in the nonrelativistic reduction the Salpeter equation does not {\it a priori} decouple with respect to $\Phi^{++}$ and $\Phi^{--}$ (see fig. (\ref{fig:rhorho3rho5})). Since distinct potential Dirac structures act on these components in different ways they should also differ in the calculation of especially higher excitation masses. This effect then can be used in order to find an appropriate Dirac structure for the confinement potential.
\subsubsection{Nonrelativistic Reduction of Model ${\cal A}$}
The special Dirac structure of model ${\cal A}$ does not mix positive ($\Phi^{++}$) and negative ($\Phi^{--}$) energy components of the Salpeter amplitude $\Phi$; the components $\Phi^{\pm\mp}$ even vanish completely as can be seen by explicit calculation:
\[ \Gamma\;\Phi\;\Gamma = \frac{1}{2}(\Id\;\Phi\;\Id - \gamma^0\;\Phi\;\gamma^0) = \left( \begin{array}{cc} 0 & \Phi^{++} \\ \Phi^{--} & 0 \end{array}\right)\; . \]
Therefore the only coupling between $\Phi^{++}$ and $\Phi^{--}$ is produced by the structure of the Salpeter equation itself due to the off-diagonal terms $\gamma^0\vec\gamma \vec p$ in the energy projectors $\Lambda^{\pm}$. In the nonrelativistic reduction, these terms vanish and the Salpeter equation with this Dirac structure then decouples with respect to $\Phi^{++}$ and $\Phi^{--}$ irrespective of calculating weakly bound states ($M\approx m_1 + m_2$), deeply bound states ($M \ll m_1 + m_2$) or higher excitation states ($M \gg m_1 + m_2$):
\[\left( \begin{array}{cc}0 & \Phi^{++}(\vec p) \\ \Phi^{--}(\vec p) & 0\end{array}\right) = \left( \begin{array}{cc}0 & \left[\int\frac{d^3 p'}{(2\pi)^3} \frac{{\cal V}_c((\vec p - \vec p\, ')^2)\;\Phi^{++}(\vec p\, ')}{M - \tilde\omega _1 - \tilde\omega _2}\right] \\ \left[\int\frac{d^3
      p'}{(2\pi)^3}\frac{{\cal V}_c((\vec p - \vec p\, ')^2)\;\Phi^{--}(\vec p\, ')}{(-M) - \tilde\omega _1 -
      \tilde\omega _2}\right] & 0\end{array}\right)\; , \]
or equivalently
\begin{eqnarray}
({\cal H}\Phi^{++})(\vec p) &=& + M\; \Phi^{++}(\vec p)\\
({\cal H}\Phi^{--})(\vec p) &=& - M\; \Phi^{--}(\vec p)\; ,
\end{eqnarray}
where the operator ${\cal H}$ is defined by:
\begin{equation}\label{eq:Hsub}
({\cal H}\Phi^{\pm\pm})(\vec p) := (\tilde \omega_1 + \tilde \omega_2)\Phi^{\pm\pm}(\vec p) + \int \frac{d^3 p\pri}{(2\pi)^3}{\cal V}_c((\vec p - \vec p\, ')^2) \Phi^{\pm\pm}(\vec p\, ')\; .
\end{equation}
For physical reasons, the potential parameters of ${\cal V}_c$ should guarantee the positive definiteness of the operator ${\cal H}$, i.e the eigenvalues $M$ of ${\cal H}$ are all positive. Therefore the negative energy component $\Phi^{--}$ has to vanish and we end up with the equation for the positive energy component $\Phi^{++}$ which is nothing else but the usual nonrelativistic Schr\"odinger equation for two spin $\frac{1}{2}$ particles moving in a spin independent confinement potential.
\subsubsection{Nonrelativistic Reduction of Model ${\cal B}$}
Whereas the structure $\frac{1}{2}(\Id\otimes\Id - \gamma^0\otimes\gamma^0)$ decouples with respect to $\Phi^{++}$ and $\Phi^{--}$, the combination $\frac{1}{2}(\Id\otimes\Id - \gamma^5\otimes\gamma^5 - \gamma^\mu\otimes\gamma_\mu)$ leads to relative strong couplings:
\[ \Gamma\;\Phi\;\Gamma = \frac{1}{2}(\Id\;\Phi\;\Id - \gamma^5\;\Phi\;\gamma^5 - \gamma^\mu\;\Phi\;\gamma_\mu) =  \left( \begin{array}{cc} 0 & \Phi^{++} \\ \Phi^{--} & 0 \end{array}\right) - \frac{1}{2}\left( \begin{array}{cc} \Phi^{-+}  & \Phi^{--} \\ \Phi^{++} & \Phi^{+-} \end{array}\right) + \frac{1}{2}\left( \begin{array}{cc} -\vec\sigma\Phi^{-+}\vec\sigma  & \vec\sigma\Phi^{--}\vec\sigma \\ \vec\sigma\Phi^{++}\vec\sigma & -\vec\sigma\Phi^{+-}\vec\sigma \end{array}\right)\; .\]
Here we have combined the timelike vector part of $\gamma^\mu\otimes\gamma_\mu$ with the scalar part such that the first term on the right hand side is nothing else but the result from the Dirac structure of model ${\cal A}$. The second term, which arises from the pseudoscalar part, swaps the positions of $\Phi^{++}$ and $\Phi^{--}$ and of $\Phi^{+-}$ and $\Phi^{-+}$, respectively. The same exchange of components is generated by the vector part but with an additional left-right multiplication by the Pauli-matrices $\vec\sigma$. 

\begin{figure}
\input{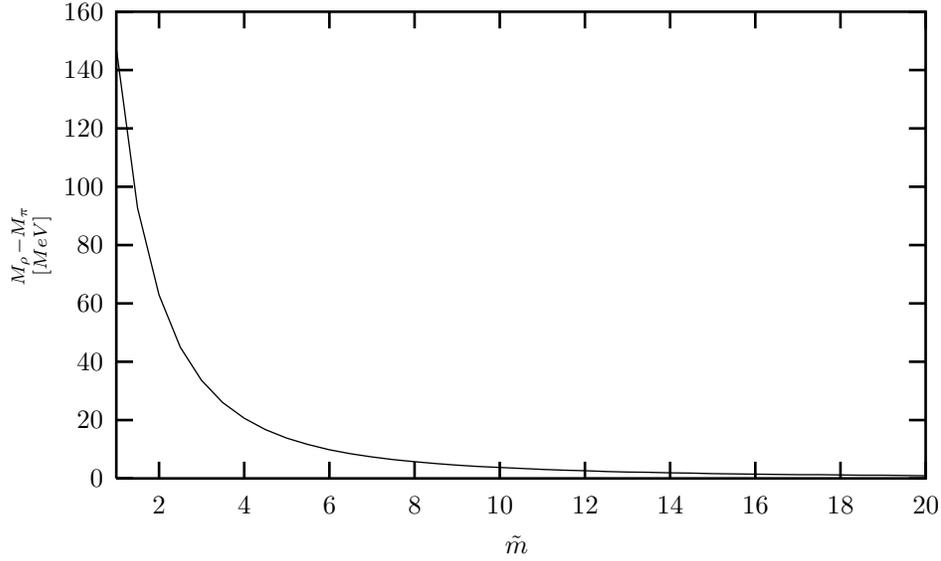}
\centering
\newline
\caption{The $\rho$-$\pi$ ground state mass splitting as a function of the mass $m_\Lambda := \tilde m \cdot m_n$ in the energy projectors $\Lambda^{\pm} =(\omega_\Lambda \pm \gamma^0(\vec \gamma 
\vec p + m_\Lambda))/2\omega_\Lambda$ with $\omega_\Lambda = \sqrt{\vec p^2 + m_\Lambda^2}$ calculated in model ${\cal B}$. In the nonrelativistic reduction ($\tilde m \longrightarrow \infty$) the (spin triplet) $\rho$ and the (spin singlet) $\pi$ are degenerate in mass due to the special combination of coefficients in the structure $\frac{1}{2}(\Id\otimes\Id - \gamma^5\otimes\gamma^5 - \gamma^\mu\otimes\gamma_\mu)$. The calculation was done without the 't Hooft interaction.}\label{fig:rhopiNONREDsp4}
\end{figure}

In the nonrelativistic reduction the components $\Phi^{+-}$ and $\Phi^{-+}$ vanish and for each spin ($S=0,1$) the Salpeter equation can be written as a system of two coupled 2$\times$2 matrix equations (see Appendix):
\begin{eqnarray}
\left[{\cal H}(\Phi^{\pm\pm} + \Phi^{\mp\mp})\right](\vec p) = \pm M \Phi^{\pm\pm}(\vec p) \;\;\;\;\;\; \mbox{for} \;\; S = 0 \label{eq:NONREDsp40}\\
\left[{\cal H}(\Phi^{\pm\pm} - \Phi^{\mp\mp})\right](\vec p) = \pm M \Phi^{\pm\pm}(\vec p) \;\;\;\;\;\; \mbox{for} \;\; S = 1 \label{eq:NONREDsp41}
\end{eqnarray}
where the operator ${\cal H}$ is defined as in eq. (\ref{eq:Hsub}).\footnote{Eq. (\ref{eq:NONREDsp40}) and eq. (\ref{eq:NONREDsp41}) exhibit an interesting symmetry concerning spin singlet and spin triplet solutions: For given mass $\pm M$ on the right hand side, a spin singlet solution ($\Phi^{++}$, $\Phi^{--}$) can be transformed into a spin triplet solution with the same mass by the transformation 
\begin{eqnarray}
(\Phi^{++}, \Phi^{--}) \longrightarrow  (\pm\Phi^{++}, \mp\Phi^{--}).
\end{eqnarray}
So, in the nonrelativistic reduction the very special combination of coefficients in the Dirac structure $\frac{1}{2}(\Id\otimes\Id - \gamma^5\otimes\gamma^5 - \gamma^\mu\otimes\gamma_\mu)$ leads to a mass degeneracy of singlet and triplet spin states although there is an explicit spin dependent term ($\vec\gamma\otimes\vec\gamma$) in the potential. As an example, fig. (\ref{fig:rhopiNONREDsp4}) shows the (spin triplet) $\rho$-mass and the (spin singlet) $\pi$-mass as a function of the parameter $\tilde m := \frac{m_{\Lambda}}{m_n}$ tuning the nonrelativistic reduction of the Salpeter equation by increasing the quark mass in the energy projectors $\Lambda^{\pm}$ indicated by $m_\Lambda$.}

\begin{figure}[b]
\input{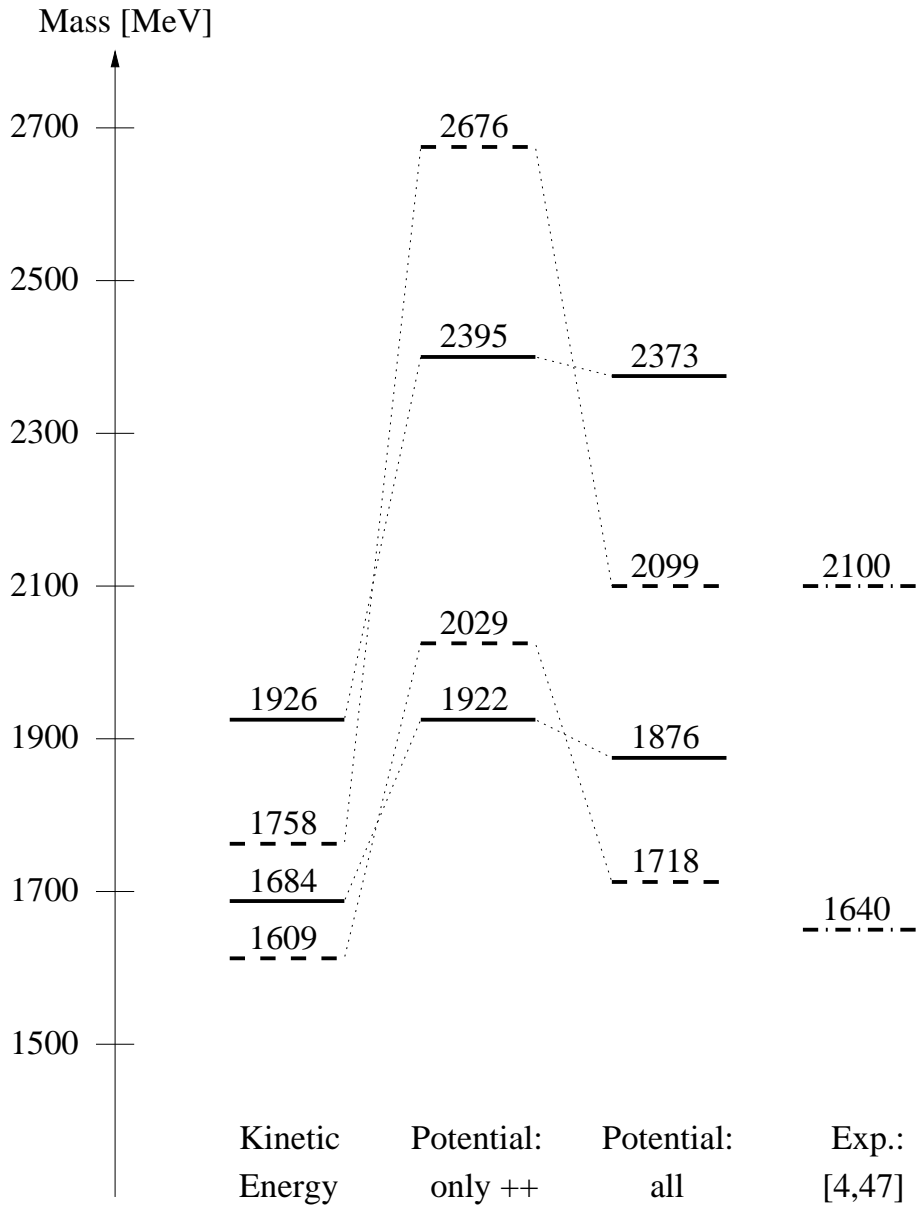}
\centering
\newline
\caption{Expectation values of the first (lower part) and second (upper part) radial $a_1$-excitation in model ${\cal A}$ (solid lines) and in model ${\cal B}$ (dashed lines); from the left to the right: kinetic energy, kinetic energy and positive (++) energy components of the potential, kinetic energy and all components of the potential (full model), experiment [4,47].}\label{fig:schspa1(1640)a1(2100)}
\end{figure}

The matrix equations eq. (\ref{eq:NONREDsp40}) and
eq. (\ref{eq:NONREDsp41}) show that even in the nonrelativistic
reduction the Dirac structure produces nonvanishing negative energy
components $\Phi^{--}$. This can be also seen in
fig. (\ref{fig:sp4_rad_a1_n0}) where the positive and negative energy
components of the ground state $a_1$ radial amplitudes are shown. The
same calculation was done with the structure
$\frac{1}{2}(\Id\otimes\Id - \gamma^0\otimes\gamma^0)$
(fig. (\ref{fig:kl_rad_a1_n0})) and as expected here the negative
energy component completely vanishes in the nonrelativistic
reduction. The fully relativistic calculation of course produces
nonvanishing negative energy components also in model ${\cal
A}$. Whereas the magnitudes of the ground state $a_1$ amplitudes in
model ${\cal A}$ and model ${\cal B}$ roughly coincide, the
corresponding amplitudes of the first radial $a_1$-excitation differ substantially 
in both models (see
fig. (\ref{fig:kl_rad_a1_n1}),
fig. (\ref{fig:sp4_rad_a1_n1})). Especially the negative energy
component produced in model ${\cal B}$
(fig. (\ref{fig:sp4_rad_a1_n1})) is much larger than its corresponding
counterpart in model ${\cal A}$ (fig. (\ref{fig:kl_rad_a1_n1})).  This
strong coupling between negative and positive energy components in
model ${\cal B}$ has of course an effect on the calculated bound state
masses. An instructive illustration of this effect is shown in
fig. (\ref{fig:schspa1(1640)a1(2100)}): Starting with the kinetic
energy we have separated the total mass of the first and second radial
$a_1$-excitation into the potential expectation values with respect to
the positive energy components only and with respect to all
components. As expected from the magnitudes of the corresponding
radial amplitudes the mass lowering caused by the negative energy
components is much larger in model ${\cal B}$ than in model ${\cal
A}$. In case of the second radial $a_1$-excitation this effect results
in a mass difference of about 300 MeV.
\subsection{Spin-Orbit Effects}
Finally, we want to point out the most outstanding feature of the Dirac structure $\frac{1}{2}(\Id\otimes\Id - \gamma^5\otimes\gamma^5 - \gamma^\mu\otimes\gamma_\mu)$, {\it i.e.} the generation of additional spin-orbit mass splittings due to the nonvanishing components $\Phi^{\pm\mp}$ of the Salpeter amplitude $\Phi$. As we will see in the next subsection, these additional spin-orbit effects in model ${\cal B}$ provide a completely different interpretation of the scalar meson ground state nonet compared to the interpretation proposed by model ${\cal A}$. In order to study the spin-orbit effects, we decompose the Dirac structure $\frac{1}{2}(\Id\otimes\Id - \gamma^5\otimes\gamma^5 - \gamma^\mu\otimes\gamma_\mu)$ with respect to the components  $\Phi^{\pm\mp}$ and introduce the spin-orbit parameter $\alpha\in[0,\frac{1}{2}]$ as follows: 
\[\Gamma\Phi\;\Gamma(\alpha) = \frac{1}{2}(\Id\;\Phi\;\Id - \gamma^0\;\Phi\;\gamma^0) - \alpha(\gamma^5\;\Phi\;\gamma^5 -\vec\gamma\;\Phi\;\vec\gamma) =  \left( \begin{array}{cc} 0 & \Phi^{++} \\ \Phi^{--} & 0 \end{array}\right) - \alpha\left( \begin{array}{cc} \Phi^{-+}  & \Phi^{--} \\ \Phi^{++} & \Phi^{+-} \end{array}\right) + \alpha\left( \begin{array}{cc} -\vec\sigma\Phi^{-+}\vec\sigma  & \vec\sigma\Phi^{--}\vec\sigma \\ \vec\sigma\Phi^{++}\vec\sigma & -\vec\sigma\Phi^{+-}\vec\sigma \end{array}\right).\]
If $\alpha = 0$, the components  $\Phi^{\pm\mp}$ vanish and the Dirac structure of model ${\cal A}$ is reproduced, {\it i.e.} $\Gamma\;\Phi\;\Gamma(0) = \frac{1}{2}(\Id\otimes\Id - \gamma^0\otimes\gamma^0)$; for $\alpha = \frac{1}{2}$, the above expression coincides with the structure of model ${\cal B}$, {\it i.e.}  $\Gamma\;\Phi\;\Gamma(\frac{1}{2}) = \frac{1}{2}(\Id\otimes\Id - \gamma^5\otimes\gamma^5 - \gamma^\mu\otimes\gamma_\mu)$. In figs. (\ref{fig:sp4_vargvg5_a1a0}) and (\ref{fig:sp4_vargvg5_K1K0st}), the ground state masses of the $a_1(1^{++})$, $a_0(0^{++})$ and the $K_1(1^+)$, $K^*_0(0^+)$ are shown as a function of the spin-orbit parameter $\alpha$. For vanishing $\alpha$, no spin-orbit effects are observed whereas for $\alpha = \frac{1}{2}$ the resulting $a_1$-$a_0$ and $K_1$-$K^*_0$ mass splitting\footnote{As the calculation was done without the 't Hooft interaction, the $\alpha$-dependence of the $f_1(1^{++})$ and $f_0(0^{++})$ completely coincides with the behaviour of their isovectorial partners $a_1(1^{++})$ and $a_0(0^{++})$.} add up to 280 MeV and 220 MeV, respectively. This splitting then yields an isovector(isoscalar) ground state $a_0$ ($f_0$) at 944 MeV and an isodoublet ground state $K^*_0$ at 1100 MeV. So, due to additional spin-orbit forces, model ${\cal B}$ produces the basic scalar ground state mass centroid at roughly 1 GeV in contrast to roughly 1.3 GeV in model ${\cal A}$. In what follows we will see that this lowering of about 300 MeV combined with the instanton induced 't Hooft interaction leads to a completely different nonet classification of the scalar mesons in both models.
\subsection{'t Hooft's Interaction and Mesons with $J = 0$}
\begin{figure}
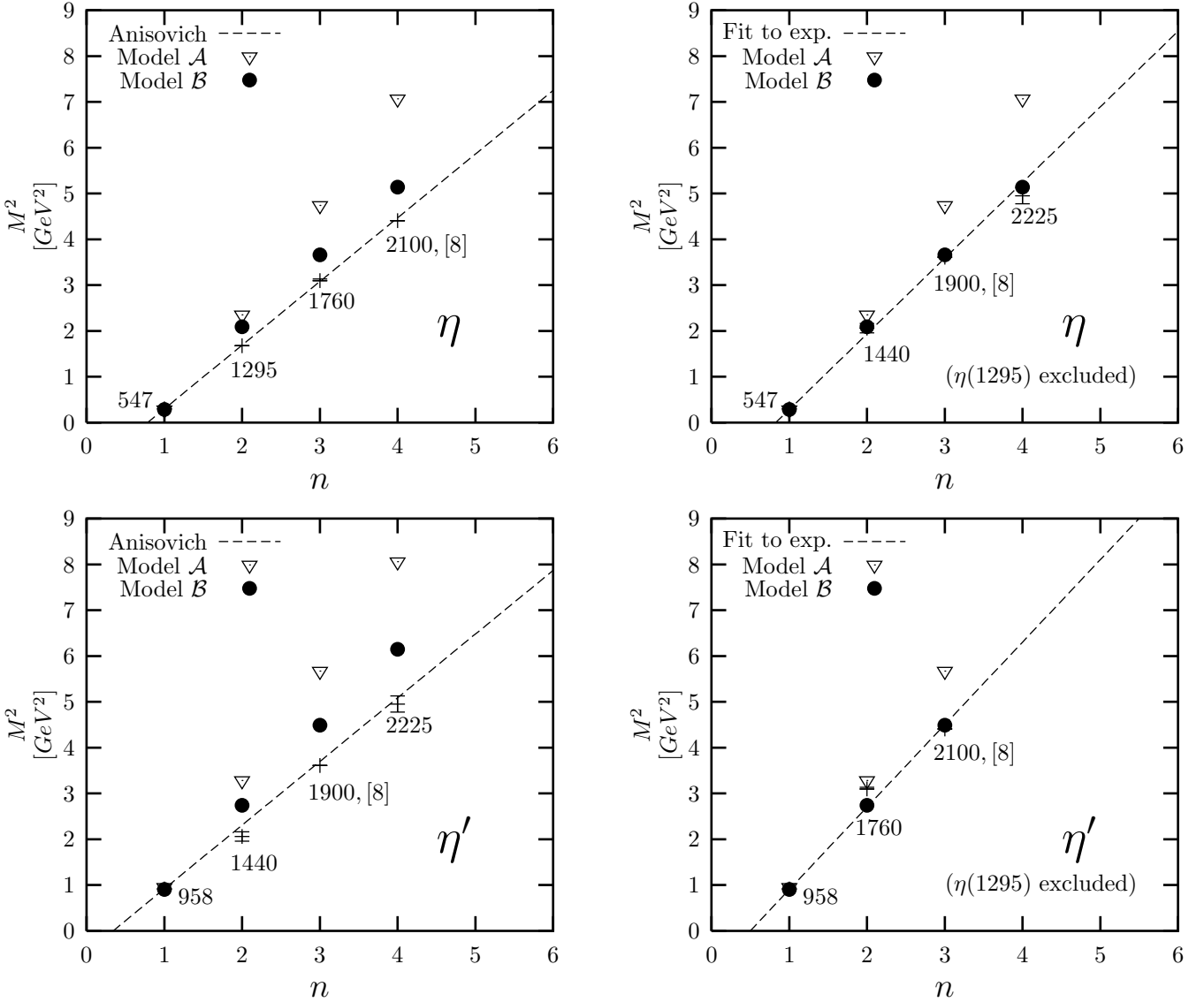

\begin{minipage}[]{17.9cm}

\begin{minipage}[t]{8.5cm}
     \input{Metavsn_new.latex}
\end{minipage}
\hspace{0.7cm}
\begin{minipage}[t]{8.5cm}
     \input{Metan1295vsn_new.latex}
\end{minipage}


\begin{minipage}[]{8.5cm}
     \input{Metapvsn_new.latex}
\end{minipage}
\hspace{0.7cm}
\begin{minipage}[]{8.5cm}
     \input{Metapn1295vsn_new.latex}
\end{minipage}

\caption{The ($n,M^2$)-trajectories for the states $\eta(00^{-+})$ and
  $\eta^{\prime}(00^{-+})$. The resonances indicated with the reference number [8] are predictions according to the formula $M^2 = M^2_0 + (n - 1) \mu^2$. The resonances without a reference number are listed in [47]. On the left hand side the trajectory slopes $\mu^2$ are taken from [8] and the $\eta(1295)$ is involved. However, the situation concerning the radial excitations of the $\eta$ and $\eta^{\prime}$ is not clear: In our model, we don't have a serious candidate for the $\eta(1295)$ to be the first radial excitation of the $\eta$. This is compatible with the observations made in [50,51]. Therefore we propose the classification of the $\eta$ and $\eta^{\prime}$ spectrum shown on the right hand side, where the $\eta(1295)$ does not occur.}\label{fig:Mvsn2}
\end{minipage}
\end{figure}

As it stands the confinement force discussed above is responsible for the global structure of the experimental mass spectrum. Especially the well established Regge trajectories can be described by this force alone. However, the comparably large mass splittings in the pseudoscalar sector require an additional residual interaction which should feature an explicit flavour dependence in order to describe the $\eta$-$\eta^\prime$-mixing and the $\eta$-$\pi$ mass difference of about 400 MeV. In fact, 't Hooft and others computed a flavour dependent effective quark interaction from instanton effects \cite{pap16,pap26,pap27} whose point-like two body part can be written as \cite{pap17}
\begin{eqnarray}
\Delta {\cal L}(2) = \frac{3}{16}\sum_i \sum_{k,l \atop{m, n}} \sum_{c_k, c_l \atop{c_m, c_n}}&g_{eff}(i)& \epsilon_{ikl}\epsilon_{imn}\left[\left(\bar\Psi_{k, c_k}\;\Id\;\Psi_{n, c_n}\right)\; \left(\bar\Psi_{l, c_l}\;\Id\;\Psi_{m, c_m}\right) + \left(\bar\Psi_{k, c_k}\;\gamma^5\;\Psi_{n, c_n}\right)\;\left(\bar\Psi_{l, c_l}\; \gamma^5 \; \Psi_{m, c_m}\right)\right]\nonumber \\
&\times& \left(\frac{3}{2}\delta_{c_k c_n}\delta_{c_l c_m} - \frac{1}{2}\delta_{c_k c_m}\delta_{c_n c_l}\right)
\end{eqnarray}
where $i,k,l,m,n$ $\in$ $\{u,d,s\}$ are flavour and $c_k, c_l, c_m, c_n \in \{r, g, b\}$ colour indices. The $\epsilon$-tensors explicitly show that this force only acts on antisymmetric flavour states. Furthermore, due to the positive sign in the Dirac structure ($\Id\otimes\Id + \gamma^5\otimes\gamma^5$), the $U_A(1)$-invariance is explicitly broken such that the $U_A(1)$-problem is in principle solved by this interaction. Note that due to the point-like nature and specific Dirac structure, the Instanton Induced Interaction (III) in the above formulation acts on states with total angular momentum $J = 0$ only. The lowest order contribution of this interaction to the Bethe-Salpeter kernel can be extracted as \cite{pap17}
\begin{eqnarray}\label{eq:THK}
\int \frac{d^3 p\pri}{(2\pi)^3} \; V_{\mbox{{\scriptsize III}}}(\vec p, \vec p\:') \Phi(\vec p\:') = 4\; G(g,g\pri)\; \int \frac{d^3p\pri}{(2\pi)^3} \; {\cal R}_\Lambda(\vec p, \vec p\:')\left(\Id \mbox{tr}\left[\Phi(\vec p\:')\right] + \gamma^5 \mbox{tr}\left[\Phi(\vec p\:') \gamma^5\right]\right)
\end{eqnarray} 
where ${\cal R}_\Lambda$ represents a regularization function and $G(g,g\pri)$ is a flavour matrix while summation over flavour indices is understood. The coupling strengths $g$ (acting on a nonstrange $\bar q q$ pair), $g^\prime$ (acting on a nonstrange-strange $\bar q q$ pair) and the finite effective range $\Lambda = \Lambda_{\mbox{{\scriptsize III}}}$ are treated as free parameters in our model. We fix them in the pseudoscalar sector in order to reproduce the experimental $\pi$-$K$-$\eta$-$\eta\pri$ mass splitting. The resulting values for the 't Hooft parameters in model ${\cal A}$ and model ${\cal B}$ can be found in table \ref{tab:Parameters} and the corresponding pseudoscalar mass spectra up to 2.5 GeV are shown in left part of fig. (\ref{fig:specj0}). For a more detailed discussion of the pseudoscalar mesons and their electroweak properties we refer to \cite{pap41}; here we only want to comment on the $\eta(1295)$ and $\eta(1440)$: both, model ${\cal A}$ and model ${\cal B}$ produce only one $\bar q q$ state in the mass region 1200--1500 MeV of the isoscalar $0^{-+}$-spectrum as can be seen in fig. (\ref{fig:specj0}). The calculated masses (1533 MeV in model ${\cal A}$, 1446 MeV in model ${\cal B}$) would prefer an identification of the $\eta(1440)$ as the first radial excitation of the ground state $\eta$ such that the $\eta(1295)$ is out of place in our quark model. Indeed, there are strong experimental hints for a non $\bar q q$ interpretation of this resonance \cite {pap50,pap51}\footnote{The author in \cite{pap50,pap51} suggest the $\eta(1440)$ to be the first radial excitation of the ground state $\eta$ compatible with our model calculations. Furthermore, in a forthcoming paper \cite{pap37} we will show, that model ${\cal B}$ predicts partial decay widths of the $\eta(1440)$ into $a_0(980)\pi$, $K^* K$ and $\eta \sigma$ in excellent agreement with the experimental values found in \cite{pap51}.} due to its absence in the reaction $p\bar p \rightarrow \pi^+\pi^-\pi^+\pi^-\eta$. Furthermore, disregarding the $\eta(1295)$ in the discussion of the $\eta$,$\eta\pri$ radial excitation spectrum, the bulk of observed and predicted isoscalar $0^{-+}$-resonances fits to the trajectory-like behaviour $M^2 = M^2_0 + (n - 1)\mu^2$ in model ${\cal B}$, with an average slope $\mu^2(00^{-+}) = 1.73 \pm 0.08$ GeV$^2$ (see the right hand side of fig. (\ref{fig:Mvsn2})).

Now, for the sake of completeness concerning the nonrelativistic reduction of our model, we also present this limit for the 't Hooft interaction:
\[\left( \begin{array}{cc}0 & \Phi^{++}(\vec p) \\ \Phi^{--}(\vec p) & 0\end{array}\right) = \left( \begin{array}{cc}0 & \left[4\; G\;\int\frac{d^3 p'}{(2\pi)^3} \frac{{\cal R}_\Lambda(\vec p, \vec p\, ')\mbox{tr}\left[\Phi^{++}(\vec p\, ') + \Phi^{--}(\vec p\, ')\right]}{M - \tilde\omega _1 - \tilde\omega _2}\right] \\ -\left[4\; G\;\int\frac{d^3
      p'}{(2\pi)^3}\frac{{\cal R}_\Lambda(\vec p, \vec p\, ')\mbox{tr}\left[\Phi^{++}(\vec p\, ') + \Phi^{--}(\vec p\, ')\right]}{M + \tilde\omega _1 +
      \tilde\omega _2}\right] & 0\end{array}\right) .\]
So, even in the nonrelativistic reduction, the 't Hooft interaction couples positive and negative energy components as can be seen from this expression. Furthermore the term proportional to the identity in eq. (\ref{eq:THK}) vanishes such that in the nonrelativistic reduction the 't Hooft interaction does not act for scalar but only for pseudoscalar mesons. In fact, the same instanton induced interaction has been used before in a nonrelativistic description within the framework of the Schr\"odinger equation\footnote{In the framework of the Schr\"odinger equation, the negative energy components $\Phi^{--}$ were completely neglected.}, both for mesons and baryons \cite{pap14}; a satisfying description of the low lying hadronic mass spectrum, especially the splitting of the pseudoscalar nonet, was obtained. However, the quoted nonrelativistic model is not able to provide a proper description of the scalar nonet due to the absence of any singlet-octet mixing mechanism in this sector. 
\subsubsection{Scalar Mesons in Model ${\cal A}$}
\begin{figure}[b]
\input{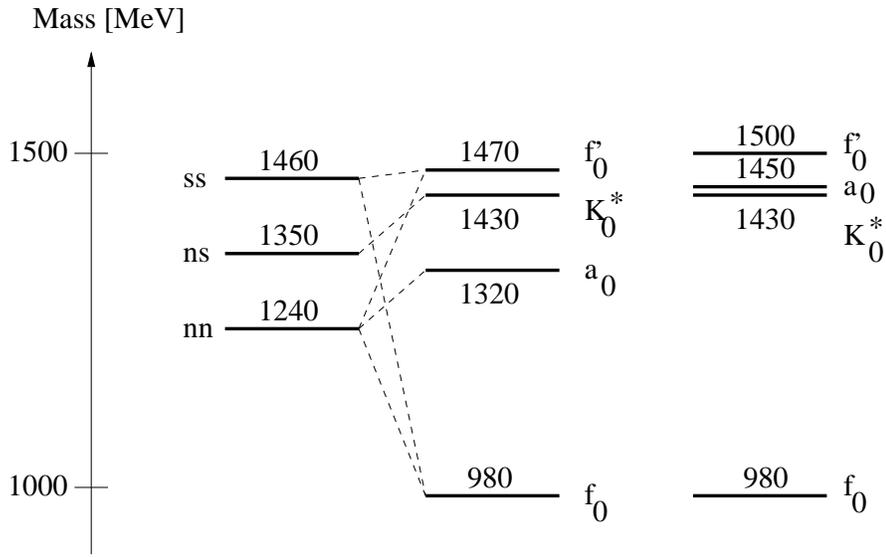}
\centering
\newline
\caption{Schematic splitting of the scalar flavour nonet with confinement model ${\cal A}$ (left), with confinement model ${\cal A}$ and instanton induced force (middle) compared to the experimental spectrum interpreted as $\bar q q$ states [47, 28, 29] (right).}\label{fig:schspscA}
\end{figure}

Now, from an earlier work \cite{pap19} we know that the present relativistic framework exhibits a natural way to cure this shortcoming. There it has been shown that the light scalar mesons can be interpreted as conventional $\bar q q$ states, with a small $SU(3)$ mixing angle, governed dynamically by 't Hooft's instanton induced interaction. The calculations were done in confinement model ${\cal A}$ such that the mass centroid of the scalar meson nonet was found at roughly 1.3 GeV as discussed in the previous subsection. Then the 't Hooft interaction, fixed in the pseudoscalar sector, provided an almost $SU(3)$ octet at about 1400 MeV and a low lying $SU(3)$ singlet at 1000 MeV. In more detail, fig. (\ref{fig:schspscA}) shows the generated mass splitting of the scalar flavour nonet due to the 't Hooft interaction in model ${\cal A}$. The dominantly singlet states are lowered whereas states with dominantly flavour octet structure are pushed to higher masses by this interaction. The calculated ground state masses are predicted to be $a_0(1321)$, $K^*_0(1426)$, $f_0(984)$ and $f_0(1468)$,
 so they imply a $\bar q q$ interpretation of the following experimental resonances (see \cite{pap47}): $a_0(1450)$, $K^*_0(1430)$, $f_0(980)$ and $f_0(1500)$. A detailed discussion of this interpretation is given in \cite{pap19}.
\subsubsection{Scalar Mesons in Model ${\cal B}$}
\begin{figure}[b]
\input{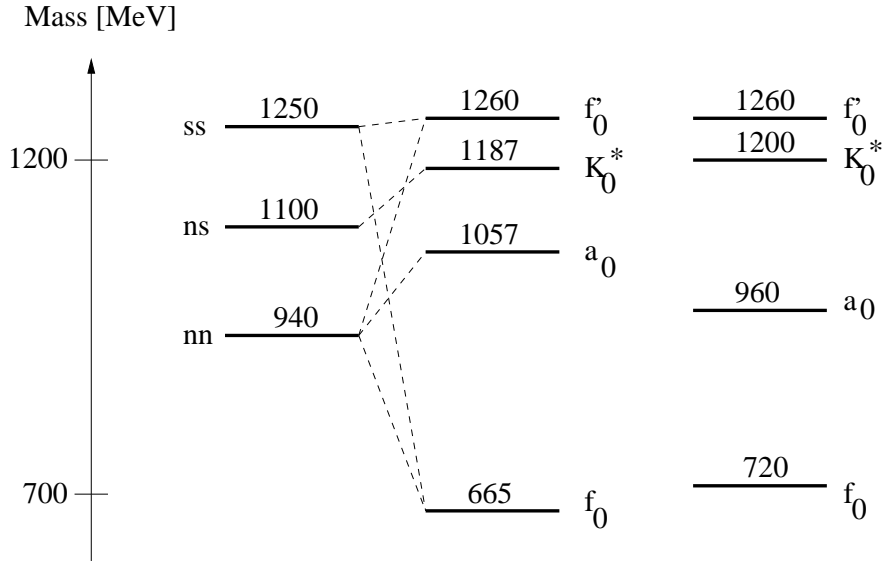}
\centering
\newline
\caption{Schematic splitting of the scalar flavour nonet with confinement model ${\cal B}$ (left), with confinement model ${\cal B}$ and instanton induced force (middle) compared to the K-matrix poles stated in [22] (right).}\label{fig:schspscB}
\end{figure}

In confinement model ${\cal B}$ the corresponding states appear at masses which are roughly 200-300 MeV lighter due to the additional spin-orbit effects in this model (see fig. (\ref{fig:schspscB})):
\begin{eqnarray}
a_0(1057), \;\;\;\;\;\; K^*_0(1187), \;\;\;\;\;\; f_0(665), \;\;\;\;\;\; f_0(1262). \nonumber  
\end{eqnarray}
Before we will compare the masses of these states not only with the resonances positions of \cite{pap47} but also with the K-matrix pole analysis of V. V. Anisovich and others (see for instance \cite{pap22}), we make some general remarks concerning the scalar nonet classification:
\begin{itemize}
\item{{\it General remarks:}}
\begin{itemize}
\item{First of all one should emphasize that, apart from the higher radial excitation states discussed in the previous subsection, nowhere else in the calculated mass spectrum the differences between model ${\cal A}$ and model ${\cal B}$ are as large as in the scalar sector discussed here. This is a very interesting observation, because the scalar sector is the only area in the experimental spectrum where mass differences of about 300 MeV do not necessarily imply a dismission of one of these models. The reason is the abundant number of experimental scalar resonances combined with many different interpretations of their physical nature (a detailed and instructive overview is given in \cite{pap30,pap31} and also in \cite{pap32}). Especially the scalar isoscalar sector features a highly complex resonance structure which to explain is still the most challenging task in meson spectroscopy. For instance, the {\sc Particle Data Group} \cite{pap47} lists the following four scalar isoscalar resonances up to 1500 MeV: $f_0(400-1200)$ (or $\sigma$-meson), $f_0(980 \pm 10)$, $f_0(1370^{+130}_{-170})$, $f_0(1500 \pm 10)$. In order to understand the internal structure of these resonances, an attempt to interpret them as members of the 1$^3$P$_0$ and 2$^3$P$_0$ light-quark nonet is of course the first step one has to do. For that, some kind of mass centroid has to be given around which the members of these nonets then split up due to different quark flavours, that is to say, due to different quark mass contents and, of course in the present model, due to different influences of the instanton induced 't Hooft interaction on these states. As the isodoublet mesons $K^*_0(1430)$ and $K^*_0(1950)$ are the least controversial of the light experimental scalar mesons and as they contain one strange and one nonstrange quark, it seems to be obvious to use their masses as fixpoints for the 1$^3$P$_0$ and 2$^3$P$_0$ quark-antiquark nonet, respectively, independent from any theoretical description. This scale then favours ground state nonet interpretations in the mass region of about 1200-1600 MeV which is in roughly agreement with the energy scale in model ${\cal A}$.\footnote{For example, Amsler and Close \cite{pap42} built a reasonable scalar nonet in the quoted mass region with the $a_0(1450)$ and $K^*_0(1430)$ setting the mass centroid, and their widths setting the scale of the nonet widths.\\
Assuming the scalar ground state nonet above 1.3 GeV, Lee and Weingarten \cite{pap53,pap54} found in their lattice gauge theory calculations the $f_0(1710)$ to be composed mainly of the lightest scalar glueball. They claim that $0^{++}$-resonances below 1.3 GeV are irrelevant to glueball spectroscopy.}}
\end{itemize}
\begin{itemize}
\item{However, V. V. Anisovich \cite{pap22} claims that the quoted mass region is noticeably higher than the average masses of other mesons which are candidates for the corresponding scalar nonet. In more detail, the experimentally well established resonances $a_0(980)$ and $f_0(980)$ are often excluded from a discussion of the scalar meson ground state nonet due to their small masses compared to the above scale. In addition, their decay properties are often entitled as unusual with respect to 'ordinary' $\bar q q$ mesons which then lead to mainly non-$\bar q q$ interpretations of these resonances\footnote{For instance, Weinstein and Isgur argue that the $a_0(980)$ and $f_0(980)$ can be understood as $\bar K K$ molecules \cite{pap33,pap38}. The mass degeneracy and their proximity to the $\bar K K$ threshold seems to require that the nature of both states must be the same. Also T. Barnes \cite{pap39} sides with these authors; he claims that the $\bar K K$ molecule picture of the $a_0(980)$ and $f_0(980)$ is supported by their small two photon decay widths. F. E. Close, Yu. L. Dokshitzer, V. N. Gribov {\it et al.} \cite{pap46} suggest to interpret the $a_0(980)$ and the $f_0(980)$ as new types of vacuum excitations ('vacuum scalars') which correspond to quark-antiquark pair creations {\it below} the Fermi surface. The J\"ulich group \cite{pap21} shows that scattering and production data on the $a_0(980)$ and $f_0(980)$ can be fitted by a sum of $s$-channel and $t$-channel exchanges without the need for genuine resonances at the $\bar K K$ threshold. On the other hand, Morgan and Pennington find a $f_0(980)$ pole structure characteristic for a genuine resonance of the constituents and not of a weakly bound system \cite{pap34,pap35}, that is to say, the extremely attractive $I = 0$ $\bar K K$ interaction may not support a loosely bound state, whereas the $I = 1$ $\bar K K$ interaction is weak and may generate a $\bar K K$ molecule, the $a_0(980)$.\\ 
However, there are also attempts to interpret both the $f_0(980)$ and the $a_0(980)$ as members of the  $\bar q q$ nonet: T\"ornqvist \cite{pap36} claims that these resonances have very large virtual components of $\bar K K$ in their wave functions. However, in order to fit the available data on the $a_0(980)$, $f_0(980)$, $f_0(1300)$ and $K^*_0(1430)$ mesons as a distorted $0^{++}$ $\bar q q$ nonet, he has to use a lot of parameters (5-6) plus an ad hoc introduced form factor which simulates the overlapping wave functions in the corresponding hadronic decay processes. Minkowski and Ochs \cite{pap52} identify the states $f_0(980)$ and $a_0(980)$ together with the $f_0(1500)$ and $K^*_0(1430)$ as the members of the scalar ground state nonet. They claim that this assignment is supported by phase shift analyses of elastic and inelastic $\pi\pi$ scattering as well as recent analyses of $\bar p p$ annihilation near threshold.}. In this context the designation 'ordinary' $\bar q q$ mesons means pure nonstrange $0^{++}\;^3$P$_0\;\bar q q$ states which could be a natural first assignment for these resonances. For instance, Godfrey and Isgur \cite{pap15} argue that this pure nonstrange classification has at least two unpleasant consequences in their 'relativized' one-gluon-exchange-plus-linear-confinement potential model: Firstly, their model does not account for the experimental fact that both states have sizeable couplings to $\bar K K$ final states, suggesting a large strange quark component. Secondly, their model provides strong decay widths $\Gamma(f_0(980) \rightarrow \pi\pi) \approx 400-1000$ MeV and $\Gamma(a_0(980) \rightarrow \eta\pi) \approx 500$ MeV which even overestimate the experimental total widths \cite{pap47} $\Gamma^{tot}_{exp}(f_0(980)) = 40-100$ MeV and $\Gamma^{tot}_{exp}(a_0(980)) = 50-100$ MeV by at least one order of magnitude. In addition, their model predicts two photon decay widths \cite{pap39} for $f_0(980)\rightarrow\gamma\gamma$ and $a_0(980)\rightarrow\gamma\gamma$ which are 5-8 times larger than the corresponding measured values. As the quoted model describes the decay properties of the vector and tensor $\bar q q$ states with satisfying success, the $f_0(980)$ and $a_0(980)$ were suggested as appropriate candidates for non-$\bar q q$ interpretations.}
\end{itemize}
\item{{\it Comparison with resonance positions of \cite{pap47}:}}
\begin{itemize}
\item{In contrast, the present fully relativistic quark model with instanton induced forces allows a $\bar q q$ interpretation of either the $f_0(980)$ (in model ${\cal A}$: the $f_0(984)$; see \cite{pap19} for a detailed interpretation of this state) or the $a_0(980)$ (in model ${\cal B}$: the $a_0(1057)$), {\it i.e.} we can not account for a $\bar q q$ interpretation of both states in a single confinement model. The model calculations of the total strong decay widths (the detailed treatment will be presented in a forthcoming paper \cite{pap37})\footnote{In \cite{pap37}, we will present the interference of two different strong decay mechanisms: instanton induced decay contributions from six quark interactions which were studied in \cite{pap40} for the first time, and quark loop contributions which not only occur for scalar and pseudoscalar mesons but also for mesons with $J > 0$.}, provide $\Gamma^{tot}(f_0(980)) = 126$ MeV and $\Gamma^{tot}(a_0(980)) = 55$ MeV in good agreement with experiment. Furthermore, the invariant coupling ratio $r(a_0(980)) = g^2_{\bar K K} / g^2_{\eta\pi}$ is predicted to be $r = 1.21$ which excellently fits to the latest K-matrix pole analysis of {\sc Crystal Barrel} and CERN-M\"unich data ($r = 1.05 - 1.35$) done by A. V. Sarantsev \cite{pap20}. Last but not least, we find $\Gamma(a_0(980)\rightarrow\gamma\gamma) = 0.50$ keV \cite{pap41} in model ${\cal B}$, which is in reasonable agreement with the experimental estimate $\Gamma^{exp}(a_0(980)\rightarrow\gamma\gamma) = (0.30 \pm 0.10)$ keV given in \cite{pap47}. In summary, the model ${\cal B}$ calculations do support a $\bar q q$ interpretation of the $a_0(980)$; even more, the results in this model may support its identification as the isovector member of the basic $1^3$ P$_0\;\bar q q$ multiplet.}
\item{In the isoscalar sector, model ${\cal B}$ yields a very low-lying dominantly singlet state at 665 MeV which may be identified with the broad structure $f_0(400-1200)$ (or $\sigma$-meson) \cite{pap47} and a state with dominantly flavour octet structure at 1262 MeV suggesting an identification with the observed $f_0(1370^{+130}_{-170})$ \cite{pap47}. Furthermore, also model ${\cal B}$ (as does model ${\cal A}$) accounts for the $f_0(1500)$ \cite{pap47} as a $\bar q q$ state\footnote{Again we refer to the paper \cite{pap37} for a more solid explanation of a $\bar q q$ interpretation of the $f_0(1500)$.} and not a glueball\footnote{The interpretation of the $f_0(1500)$ as the ground state glueball mixed with close-by conventional scalar mesons is strongly favoured by Amsler and Close \cite{pap42}.} as can be seen in the right part of fig. (\ref{fig:specj0}) where the complete scalar excitation spectrum up to 2.5 GeV is shown. Whereas in model ${\cal A}$ the calculated mass at 1468 MeV is identified with the dominantly octet state of the scalar ground state nonet, model ${\cal B}$ provides a dominantly singlet state at 1554 MeV which then belongs to the first excited scalar nonet. However, in view of the complexity in this sector that arises from strong decay channel couplings and possible mixtures with gluonic or other exotic states, one has to regard also decay observables for a more realistic interpretation of the scalar mesons. Some results concerning the $\gamma\gamma$ decays of these states can be found in \cite{pap41}; a more detailed discussion which also includes numerical results on strong decay widths will be given in a forthcoming paper (see \cite{pap37}).}
\item{The classification of the calculated scalar isodoublet states with respect to the listed $\frac{1}{2}0^+$-resonances of the {\sc Particle Data Group} \cite{pap47} faces a problem in model ${\cal B}$: compared to the PDG value of the lightest scalar kaon, $M_{K^*_0} \approx 1430$ MeV (see \cite{pap47}), the corresponding model ground state appears at 1187 MeV. As discussed above, this lowering of about 250 MeV is a direct consequence from the additional spin-orbit forces in confinement model ${\cal B}$ (see fig. (\ref{fig:sp4_vargvg5_K1K0st})).}
\end{itemize}
\item{{\it Comparison with K-matrix poles from Anisovich et al.:}}
\begin{itemize}
\item{V. V. Anisovich and coworkers \cite{pap22,pap23,pap24,pap43,pap44,pap45} suggested to identify the $\bar q q$ states not with mean resonance positions but rather with the poles of the K-matrix fits to the relevant data sets, the so-called 'bare states'. They emphasize that pure quark model calculations (as for instance the present one) do not take into account the resonance decay, that is to say, these calculations neglect any effects of decay-channel couplings on the meson masses. Therefore there are good reasons to compare our calculated masses to bare states. Unfortunately, up to now such a K-matrix analysis has not been performed for all quantum numbers, so in many meson sectors the bare states are unknown such that an overall comparison with our masses is not possible. Moreover, from our point of view in many sectors significant differences in mass shifts between bare states and real resonances would be unpleasant due to the fact that our model calculations agree well with the global structure of the experimental mass spectrum.}
\item{However, stimulated by the problems mentioned above with respect to the scalar nonet classification just in this sector a multitude of experimental data sets has been analysed by K-matrix pole techniques. The most interesting result of these analyses is, that the members of the bare scalar ground state nonet do not appear around 1.3-1.4 GeV but in a mass region which is roughly 200-300 MeV lighter than this scale, namely \cite{pap22,pap23,pap24,pap43,pap44,pap45}
\begin{eqnarray}
a^{bare}_0(960\pm 30), \;\;\;\;\;\; K^{*\; bare}_0(1200^{+90}_{-150}), \;\;\;\;\;\; f^{bare}_0(720\pm 100), \;\;\;\;\;\; f^{bare}_0(1260^{+100}_{-30}). \nonumber  
\end{eqnarray}
In particular, the lightest scalar bare kaon appears 200 MeV lower than the amplitude pole. And in fact, the calculated scalar ground state nonet of model ${\cal B}$ shows a remarkable agreement with these bare states as can be seen in fig. (\ref{fig:schspscB}).\footnote{The low absolute value for the scalar mass centroid of about 1.0--1.1 GeV in model ${\cal B}$ is not really astonishing due to the $U_A(1)$-invariance of the Dirac structure $\frac{1}{2}(\Id\otimes\Id-\gamma^5\otimes\gamma^5-\gamma^\mu\otimes\gamma_\mu):$ As discussed above, in the limit of vanishing constituent quark masses this invariance leads to parity doublets in the meson mass spectrum. So, in this model the mass splitting between states which only differ in their parity quantum number is mainly caused by the nonvanishing quark mass terms in the Salpeter equation. The ground states of the strange vector mesons $K^*$ and $K_1$ then provide an appropriate estimate of this mass splitting. Their calculated masses differ by approximately $\Delta M \approx 400$ MeV which also fits to the splitting between the ground state masses of the nonstrange parity partners $\rho$ and $a_1$. Finally, neglecting the $U_A(1)$-violating 't Hooft interaction the mass of the lightest pseudoscalar kaon $K(0^-)$ would appear at roughly 700 MeV (see \cite{pap41}) indicating to expect the ground state mass of the parity partner $K^*_0(0^+)$ around 700 MeV $+$ 400 MeV $=$ 1100 MeV, and in fact, this is the value produced by the model ${\cal B}$ calculations (see fig. (\ref{fig:schspscB})).}}
\item{In addition to the scalar ground state nonet, the scalar first excited nonet of confinement model ${\cal B}$
\begin{eqnarray}
a_0(1665), \;\;\;\;\;\; K^*_0(1788), \;\;\;\;\;\; f_0(1554), \;\;\;\;\;\; f_0(1870) \nonumber  
\end{eqnarray}
also coincides with the K-matrix results of the authors in \cite{pap22,pap23,pap24,pap43,pap44,pap45}. They found the following bare $2^3$P$_0\;\bar q q$ nonet\footnote{There exists a second K-matrix solution for the $2^3$P$_0\;\bar q q$ nonet: the authors in \cite{pap22,pap23,pap24,pap43,pap44,pap45} find a third scalar isoscalar bare state between 1200-1600 MeV, the $f^{bare}_0(1235\pm 50)$, which, instead of the $f^{bare}_0(1600 \pm 50)$, can also be used to complete the $2^3$P$_0\; \bar q q$ nonet. So, one of the scalar isoscalar bare states in the mass region 1200-1600 MeV is superfluous for the $\bar q q$ classification and may be connected to the lightest scalar glueball. The K-matrix solutions lead to positions of the amplitude poles in the complex mass plane whose real part may be compared to average mass values taken from the {\sc Particle Data Group} \cite{pap47} ([position of the real part in MeV, PDG 00 \cite{pap47}]): $[988\pm 6, a_0(980)], [1415\pm 25, K^*_0(1430)], [1015\pm 15, f_0(980)], [1300\pm 20, f_0(1370^{+130}_{-170})], [1499\pm 8, f_0(1500)], [1530^{+90}_{-250}, f_0(1370^{+130}_{-170})], [1565\pm 30, a_0(1450)], [1780\pm 30, f_0(1710)], [1820\pm 40, K^*_0(1950)]$.}:
\begin{eqnarray}
a^{bare}_0(1640\pm 40), \;\;\;\;\;\; K^{*\; bare}_0(1820^{+40}_{-60}), \;\;\;\;\;\; f^{bare}_0(1600\pm 50), \;\;\;\;\;\; f^{bare}_0(1810^{+30}_{-100}). \nonumber 
\end{eqnarray}
}
\end{itemize}
\end{itemize}
Summarizing, the present fully relativistic confining quark model with 't Hooft's instanton induced force as residual interaction allows to generate scalar $\bar q q$ states, whose masses and flavour mixings essentially depend on the confinement Dirac structure: The $U_A(1)$-violating structure $\frac{1}{2}(\Id\otimes\Id - \gamma^0\otimes\gamma^0)$ fixes the basic mass centroid of the ground state nonet at roughly 1.3 GeV, which the 't Hooft interaction then splits into an almost pure flavour singlet $f_0$-state at roughly 1 GeV and the flavour octet states $f_0$, $a_0$, $K^*_0$ in the mass region around 1.4 GeV (see \cite{pap19} for a detailed interpretation of these states).\\
In contrast, the $U_A(1)$-invariant Dirac structure $\frac{1}{2}(\Id\otimes\Id - \gamma^5\otimes\gamma^5 - \gamma^\mu\otimes\gamma_\mu)$ produces the basic mass centroid of the scalar ground state nonet at about 1.0-1.1 GeV. Here the 't Hooft interaction lowers the dominantly flavour singlet $f_0$-state to roughly 700 MeV and pushes the dominantly flavour octet $f_0$-state to approximately 1.3 GeV. The lightest $a_0$-state appears at 1057 MeV and its decay properties are compatible with a $\bar q q$ interpretation of the $a_0(980)$. The lightest scalar kaon appears roughly 200 MeV lower than the corresponding PDG-resonance $K^*_0(1430)$ \cite{pap47}. However, as the model presented here does not take into account any effects of decay-channel couplings on the meson masses, we followed a suggestion of V. V. Anisovich and coworkers \cite{pap22,pap23,pap24,pap43,pap44,pap45}, not to identify our calculations with the real observed resonances but with the K-matrix poles deduced from appropriate data sets. This assignment then not only fits to the calculated members of the ground state scalar nonet but also to their first radial excitation states in confinement model ${\cal B}$. 

\section{Summary and Conclusion}
Within the framework of a relativistic quark model based on the instantaneous Bethe-Salpeter equation, we have studied two different Dirac structures for a linearly rising confinement potential with respect to the complete meson spectrum. It was shown that an $U_A(1)$-invariant structure of the form $\frac{1}{2}(\Id\otimes\Id - \gamma^5\otimes\gamma^5 - \gamma^\mu\otimes\gamma_\mu)$ (model ${\cal B}$) provides an excellent description of the experimentally well established Regge trajectories just as the scalar time-like vector combination $\frac{1}{2}(\Id\otimes\Id - \gamma^0\otimes\gamma^0)$ (model ${\cal A}$) which has been used in earlier works. 
However, since the corresponding radial excitation spectra were found to be very different in both models we extended the comparison with the experimental data to a multitude of new resonances in the mass region 1000--2400 MeV observed in the {\sc Crystal Barrel} $\bar p N$-annihilation data during the last few years. Whereas the model calculations done with the structure $\frac{1}{2}(\Id\otimes\Id - \gamma^0\otimes\gamma^0)$ overestimate these new data by roughly 150--350 MeV, the combination  $\frac{1}{2}(\Id\otimes\Id - \gamma^5\otimes\gamma^5 - \gamma^\mu\otimes\gamma_\mu)$ produces masses in remarkably good agreement with the newly observed resonances. Here the deviation is in general less than 100 MeV. Furthermore, the squared masses of the excited states in model ${\cal B}$ show for given quantum numbers, in contrast to model ${\cal A}$, a linear dependence on their radial excitation number, $M^2 \propto n$, very similar to the behaviour of many experimental resonances recently observed by A. V. Anisovich, V. V. Anisovich and A. V. Sarantsev \cite{pap08}. We thus can now relate this observation to the quark-antiquark confinement dynamics in the framework of a constituent quark model. We found that the lowering of radial excited states in model ${\cal B}$ is a direct consequence of the strong coupling between positive and negative Salpeter energy components due to the $\gamma^5$- and $\gamma^\mu$-part in the structure $\frac{1}{2}(\Id\otimes\Id - \gamma^5\otimes\gamma^5 - \gamma^\mu\otimes\gamma_\mu)$. This effect was illustrated by carrying out the nonrelativistic reduction of both Dirac structures and it turned out that the Salpeter equation in model ${\cal A}$ reduces to the usual Schr\"odinger equation whereas in model ${\cal B}$ the negative energy components do not vanish even in this limit. 
By virtue of these observations one can conclude that any quark model should account for the correct treatment of relativistic effects not only in the calculations of deeply bound state masses ($M \ll m_1 + m_2$) but also in the case of higher radial excitation states ($M \gg m_1 + m_2$). Therefore the mass spectra calculations performed in the fully relativistic Salpeter framework as presented in this paper, may be more realistic than the numerous so-called 'relativized' quark model calculations that do not incorporate the negative energy components of the constituents in an unrestricted manner.

A further feature of the fully relativistic Salpeter framework affects the scalar sector: Using the $U_A(1)$-breaking instanton induced 't Hooft interaction in order to compute the pseudoscalar mass splittings, the fully relativistic framework automatically generates flavour mixings also in the scalar sector caused by the same interaction. Therefore the present model allows to compute realistic scalar $\bar q q$ states, in contrast to its nonrelativistic reduction, where the 't Hooft interaction acts for the pseudoscalar mesons only. It turned out that the masses and flavour mixings of the scalar mesons essentially depend on the confinement Dirac structure: In model ${\cal A}$ the structure $\frac{1}{2}(\Id\otimes\Id - \gamma^0\otimes\gamma^0)$ fixes the basic mass centroid of the scalar ground state nonet at roughly 1.3 GeV, which the 't Hooft interaction then splits into an almost $SU(3)$ singlet at roughly 1 GeV and an almost $SU(3)$ octet at about 1.4 GeV. In contrast, the $U_A(1)$-invariant Dirac structure  $\frac{1}{2}(\Id\otimes\Id - \gamma^5\otimes\gamma^5 - \gamma^\mu\otimes\gamma_\mu)$ in model ${\cal B}$ produces the basic mass centroid at about 1.0--1.1 GeV. Here the 't Hooft interaction lowers the dominantly flavour singlet $f_0$-state to roughly 700 MeV and pushes the dominantly flavour octet $f_0$-state to approximately 1.3 GeV. The lightest $a_0$-state appears at roughly 1 GeV and its decay properties are not in contradiction with a $\bar q q$ interpretation of the $a_0(980)$. The lightest scalar kaon appears roughly 200 MeV lower than the corresponding experimental resonance in this sector. However, as the model presented here does not take into account any effects of decay channel couplings on the meson masses, we followed a suggestion of V. V. Anisovich and coworkers, not to identify our calculations with the observed resonance positions but with their K-matrix poles ('bare states') deduced from appropriate data sets. This assignment then not only fits to the calculated members of the scalar ground state nonet but also to their first radial excitation states in confinement model ${\cal B}$. However, up to now such a K-matrix analysis has not been performed for all quantum numbers. Therefore the 'bare states' are unknown in many meson sectors, such that an overall comparison with our masses is not possible so far. Moreover, from our point of view significantly different mass shifts between 'bare states' and real resonances would be unpleasant in many sectors due to the fact that our model calculations agree well with the global structure of the experimental mass spectrum. A K-matrix analysis in all meson sectors would be very desirable to clarify this matter.  

\section*{Appendix: Nonrelativistic Reduction of  Model ${\cal B}$}
The action of the Dirac structure $\frac{1}{2}(\Id\otimes\Id - \gamma^5\otimes\gamma^5 - \gamma^\mu\otimes\gamma_\mu)$ on the Salpeter amplitude 
\[\Phi = \left( \begin{array}{lr}\Phi^{+-} & \Phi^{++} \\ \Phi^{--} & \Phi^{-+}\end{array}\right) \]
can be written in the form
\[ \Gamma\;\Phi\;\Gamma = \frac{1}{2}(\Id\;\Phi\;\Id - \gamma^5\;\Phi\;\gamma^5 - \gamma^\mu\;\Phi\;\gamma_\mu) =  \left( \begin{array}{cc} 0 & \Phi^{++} \\ \Phi^{--} & 0 \end{array}\right) - \frac{1}{2}\left( \begin{array}{cc} \Phi^{-+}  & \Phi^{--} \\ \Phi^{++} & \Phi^{+-} \end{array}\right) + \frac{1}{2}\left( \begin{array}{cc} -\vec\sigma\Phi^{-+}\vec\sigma  & \vec\sigma\Phi^{--}\vec\sigma \\ \vec\sigma\Phi^{++}\vec\sigma & -\vec\sigma\Phi^{+-}\vec\sigma \end{array}\right).\]
For a better understanding of the explicit spin dependence, it is useful to rewrite the left-right multiplication $\vec\sigma\otimes\vec\sigma$ into a spin projector multiplication from the left:
\begin{eqnarray}
\vec\sigma\Phi^{ij}\vec\sigma = \left(3\; {\cal P}_{S=0} - {\cal P}_{S=1}\right)\Phi^{ij}, \;\;\;\; i,j \in \{+,-\}
\end{eqnarray} 
where ${\cal P}_{S=0}$ and  ${\cal P}_{S=1}$ denote the projectors on spin singlet ($S=0$) and spin triplet ($S=1$) states with ${\cal P}_{S=0} + {\cal P}_{S=1} = \Id_S$ where $\Id_S$ is the identity on the spin space. The left-right multiplication $\vec\gamma\otimes\vec\gamma$ then can be written as 
\[ \vec\gamma\;\Phi\;\vec\gamma = \left(3\;{\cal P}_{S=0} - {\cal P}_{S=1}\right)\left( \begin{array}{cc} -\Phi^{-+}  & \Phi^{--} \\ \Phi^{++} & -\Phi^{+-} \end{array}\right)\]
such that we end up with 
\[ \Gamma \Phi\;\Gamma = \frac{1}{2}(\Id\otimes\Id - \gamma^5\otimes\gamma^5 - \gamma^\mu\otimes\gamma_\mu) = \\
{\cal P}_{S=0}\left( \begin{array}{cc} -2\;\Phi^{-+}  &  \Phi^{++} + \Phi^{--} \\ \Phi^{--} + \Phi^{++} & -2\;\Phi^{+-} \end{array}\right) + {\cal P}_{S=1}\left( \begin{array}{cc} 0  &  \Phi^{++} - \Phi^{--} \\ \Phi^{--} - \Phi^{++} & 0 \end{array}\right). \]
In the nonrelativistic reduction the components $\Phi^{+-}$ and $\Phi^{-+}$ vanish and for both spins the Salpeter equation can be written as a system of two coupled 2$\times$2 matrix equations:
\begin{eqnarray}
\left[{\cal H}(\Phi^{\pm\pm} + \Phi^{\mp\mp})\right](\vec p) = \pm M \Phi^{\pm\pm}(\vec p) \;\;\;\;\;\; \mbox{for} \;\; S = 0 \label{APPCeq:NONREDsp40}\\
\left[{\cal H}(\Phi^{\pm\pm} - \Phi^{\mp\mp})\right](\vec p) = \pm M \Phi^{\pm\pm}(\vec p) \;\;\;\;\;\; \mbox{for} \;\; S = 1 \label{APPCeq:NONREDsp41}
\end{eqnarray}
where the operator ${\cal H}$ is defined as in eq. (\ref{eq:Hsub}).
\section*{Figures}
\begin{figure}[t]
\input{klempt_sp4_T0_new.pstex_t}
\centering
\newline
\caption{Isoscalar meson spectrum ($J > 0$). In the middle of each column the experimental resonance positions [47] and their errors are indicated by lines and shaded areas; the lines in the left and right part of each column represent the calculated masses in model ${\cal A}$ and in model ${\cal B}$, respectively.}
\label{fig:spect0}
\end{figure}
\begin{figure}[t]
\input{klempt_sp4_T1_new.pstex_t}
\centering
\newline
\caption{Isovector meson spectrum ($J > 0$). In the middle of each column the experimental resonance positions [47] and their errors are indicated by lines and shadowed areas; the lines in the left and right part of each column represent the calculated masses in model ${\cal A}$ and model ${\cal B}$, respectively.} 
\label{fig:spect1}
\end{figure}
\newpage

\vspace*{-2.30cm}

\begin{figure}
\begin{minipage}[]{17.9cm}
\begin{minipage}[t]{8.5cm}
     \input{Mpivsn_new.latex}
\end{minipage}
\hspace{0.7cm}
\begin{minipage}[t]{8.5cm}
     \input{Ma0vsn_new.latex}
\end{minipage}


\begin{minipage}[]{8.5cm}
     \input{MrhoSdomvsn_new.latex}
\end{minipage}
\hspace{0.7cm}
\begin{minipage}[]{8.5cm}
     \input{Ma1vsn_new.latex}
\end{minipage}

\begin{minipage}[b]{8.5cm}
     \input{Ma2Pdomvsn_new.latex}
\end{minipage}
\hspace{0.7cm}
\begin{minipage}[b]{8.5cm}
     \input{Mpi2vsn_new.latex}
\end{minipage}
\caption{The ($n,M^2$)-trajectories for the states $\pi(10^{-+})$,
  $a_0(10^{++}), \rho(11^{--}), a_1(11^{++}), a_2(12^{++})$ and $\pi_2(12^{-+})$. The resonances indicated with the reference number [8] are predictions according to the formula $M^2 = M^2_0 + (n - 1) \mu^2$. The trajectory slopes $\mu^2$ are taken from [8] and are approximately the same for all
  trajectories. The resonances without a reference number are well established and listed in [47]. All other resonances have been indicated with the reference number where they have been seen or predicted. The trajectory-like behaviour found in model ${\cal B}$ fits remarkably good to the slopes stated in [8].}\label{fig:Mvsn1}
\end{minipage}
\end{figure}

\newpage
\vspace*{-1.95cm}

\newpage

\newpage
\begin{figure}
\begin{minipage}[]{17.9cm}

\begin{minipage}[t]{8.5cm}
\input{nonrel_kl_a1_n0_radial.latex}
\end{minipage}
\hspace{0.7cm}
\begin{minipage}[t]{8.5cm}
\input{kl_a1_n0_radial.latex}
\end{minipage}
\vspace*{0.25cm}
\caption{Positive (${\cal R}^{+}$) and negative (${\cal R}^{-}$) energy components of the ground state radial $a_1$-amplitude in the nonrelativistic (left) and fully relativistic version (right) of model ${\cal A}$.}\label{fig:kl_rad_a1_n0}
\end{minipage}
\end{figure}

\vspace*{3cm}

\begin{figure}
\begin{minipage}[]{17.9cm}

\begin{minipage}[]{8.5cm}
\input{nonrel_sp4_a1_n0_radial.latex}
\end{minipage}
\hspace{0.7cm}
\begin{minipage}[]{8.5cm}
\input{sp4_a1_n0_radial.latex}
\end{minipage}
\vspace*{0.25cm}
\caption{Positive (${\cal R}^{+}$) and negative (${\cal R}^{-}$) energy components of the ground state radial $a_1$-amplitude in the nonrelativistic (left) and fully relativistic version (right) of model ${\cal B}$.}\label{fig:sp4_rad_a1_n0}
\end{minipage}
\end{figure}

\newpage

\begin{figure}
\begin{minipage}[]{17.9cm}

\begin{minipage}[t]{8.5cm}
\input{nonrel_kl_a1_n1_radial.latex}
\end{minipage}
\hspace{0.7cm}
\begin{minipage}[t]{8.5cm}
\input{kl_a1_n1_radial_new.latex}
\end{minipage}
\vspace*{0.25cm}
\caption{Positive (${\cal R}^{+}$) and negative (${\cal R}^{-}$) energy components of the first excited radial $a_1$-amplitude in the nonrelativistic (left) and fully relativistic version (right) of model ${\cal A}$.}\label{fig:kl_rad_a1_n1}
\end{minipage}
\end{figure}

\vspace*{3cm}

\begin{figure}
\begin{minipage}[]{17.9cm}

\begin{minipage}[]{8.5cm}
\input{nonrel_sp4_a1_n1_radial.latex}
\end{minipage}
\hspace{0.7cm}
\begin{minipage}[]{8.5cm}
\input{sp4_a1_n1_radial_new.latex}
\end{minipage}
\vspace*{0.25cm}
\caption{Positive (${\cal R}^{+}$) and negative (${\cal R}^{-}$) energy components of the first excited radial $a_1$-amplitude in the nonrelativistic (left) and fully relativistic version (right) of model ${\cal B}$.}\label{fig:sp4_rad_a1_n1}
\end{minipage}
\end{figure}

\newpage

\newpage

\begin{figure}
\input{sp4_vargvg5_a1a0.latex}
\centering
\newline
\caption{The ground state $a_1(1^{++})$- and $a_0(0^{++})$-mass as a function of the spin-orbit parameter $\alpha$ calculated with the confinement parameters of model ${\cal B}$ (A = $-1135$ MeV, B = 1300 MeV/fm). For $\alpha = 0$, the Dirac structure $\frac{1}{2}(\Id\otimes\Id - \gamma^0\otimes\gamma^0) - \alpha (\gamma^5\otimes\gamma^5 - \vec\gamma\otimes\vec\gamma)$ reduces to the structure of model ${\cal A}$ and no spin-orbit splitting is observed; for $\alpha = 0.5$, it coincides with the structure of model ${\cal B}$ and generates a spin-orbit splitting of about 280 MeV. As the calculation was done without the 't Hooft interaction, the isoscalar partners $f_1(1^{++})$ and $f_0(0^{++})$ show the same behaviour with respect to the spin-orbit parameter $\alpha$.}\label{fig:sp4_vargvg5_a1a0}
\end{figure}
\begin{figure}
\input{sp4_vargvg5_K1K0st.latex}
\centering
\newline
\caption{The ground state $K_1(1^{+})$- and $K^{*}_0(0^{+})$-mass as a function of the spin-orbit parameter $\alpha$ calculated with the parameters of model ${\cal B}$ (A = $-1135$ MeV, B = 1300 MeV/fm). For $\alpha = 0$, the Dirac structure $\frac{1}{2}(\Id\otimes\Id - \gamma^0\otimes\gamma^0) - \alpha (\gamma^5\otimes\gamma^5 - \vec\gamma\otimes\vec\gamma)$ reduces to the structure of model ${\cal A}$ and no spin-orbit splitting is observed; for $\alpha = 0.5$, it coincides with the structure of model ${\cal B}$ and generates a spin-orbit splitting of about 220 MeV. The calculation was done without the 't Hooft interaction.}\label{fig:sp4_vargvg5_K1K0st}
\end{figure}

\newpage

\begin{figure}[t]
\input{klempt_sp4_J0_new.pstex_t}
\centering
\newline
\caption{Pseudoscalar and scalar meson spectrum. In the middle of each column the experimental resonance positions [47] and their errors are indicated by lines and shaded areas; the lines in the left and right part of each column represent the calculated masses in model ${\cal A}$ and model ${\cal B}$, respectively.} 
\label{fig:specj0}
\end{figure}
\newpage

\newpage

\vspace*{1.5cm}

\newpage

\section*{Tables}
\begin{table}
\caption{The parameters of the confinement force, the 't
Hooft interaction and the constituent quark masses in the models
$\cal A$ and $\cal B$.
}
\label{tab:Parameters}
  \begin{tabular}{cccc}
\protect
\rule[-6mm]{0mm}{13mm} & {\bf Parameter} & {\bf Model $\cal A$ } & {\bf
Model $\cal B$} \\
\hline 
 & & & \\
& $g$ [GeV${}^{-2}$] 		& 1.73 & 1.62  \\
\raisebox{2ex}[1ex]{\sl 't Hooft} &$g'$ [GeV${}^{-2}$] 	& 1.54 & 1.35  \\ 
\raisebox{2ex}[1ex]{\sl interaction}& $\Lambda _{\mbox{\scriptsize III}}$ [fm] & 0.30 & 0.42\\ 
 & & & \\
{\sl Constituent}& $m_n$ [MeV] & 306 & 380 \\
{\sl quark masses}&$m_s$ [MeV] & 503 & 550 \\
 & & & \\
 {\sl Confinement}& $a_c$ [MeV] & --1751 & --1135 \\
 {\sl parameters}& $b_c$ [MeV/fm] & 2076 & 1300 \\
& & & \\
{\sl Spin structure} &${\Gamma\otimes\Gamma}$ & $\frac 1 2 (\Id\otimes\Id - \gamma^0\otimes\gamma^0)$ &
$\frac 1 2 (\Id\otimes\Id - \gamma^5\otimes\gamma ^5 - \gamma
^\mu\otimes\gamma _\mu)$  \\
 & & & \\
  \end{tabular}
\end{table}


\begin{table}
\begin{center}
\caption{New candidates for isovector resonances and their masses in [MeV]. The first and second columns show the name and publication reference of each resonance, respectively; the third column contains the experimental mass and its error; the calculated masses in the full Salpeter model (FSM) are shown in the fourth (model ${\cal B}$) and in the seventh (model ${\cal A}$) column. The corresponding results in their nonrelativistic reduction models (NRM) are also shown. The sixth column shows some calculations done by Godfrey and Isgur in their relativized quark model [15]. The dominantly P-wave and F-wave states of the $a_2$ are labeled by P and F, respectively. The dominantly F-wave and H-wave states of the $a_4$ are labeled by F and H, respectively.}
\label{tab1}
\begin{tabular}{cccccccc}
&&&&&&&\\
&&& \multicolumn{2}{c}{{\bf Model ${\cal B}$}} && \multicolumn{2}{c}{{\bf Model ${\cal A}$}} \\
&&&&&&&\\
{\bf Name} & {\bf Ref.} & {\bf Exp. Mass} & {\bf FSM} & {\bf NRM} & {\bf Godfrey \& Isgur} & {\bf FSM} & {\bf NRM} \\
&&&&&&&\\
\cline{1-8}
&&&&&&&\\
$a_0(2025)$ & [1] & 2025 $\pm$ 30 & 2071 & 2250 & -- & 1932\tablenotemark[1] & 1927\tablenotemark[1] \\
&&&&&&&\\
&&&&&&&\\
$a_1(1640)$ & [47] & 1640 $\pm$ 42 &&&&& \\
&&& 1718 & 1845 & 1820 & 1876  & 1927 \\
$a_1(1700)$ & [3] & 1700 &&&&& \\
&&&&&&&\\
&&&&&&&\\
$a_1(2100)$ & [4] & 2100 $\pm$ 20 & 2099 & 2250 & -- & 2374 & 2528 \\
&&&&&&&\\
$a_1(2340)$ & [4] & 2340 $\pm$ 40 & 2412 & 2586 & -- & 2791 & 3073 \\
&&&&&&&\\
&&&&&&&\\
$a_2(1660)$ & [47] & 1660 $\pm$ 40 &&&&& \\
&&& P:1807 & P:1845 & P:1820 & P:1931 & P:1927 \\
&&& F:1768 & F:1986 & $\;\;$F:2050\tablenotemark[3] & F:1879 & F:2002 \\
$a_2(1750)$ & [47] & 1752 $\pm$ 25 &&&&& \\
&&&&&&&\\
&&&&&&&\\
$a_2(2100)$ & [4] & 2100$^{+10}_{-30}$ & P:2160 & P:2250 & -- & P:2411 & P:2528 \\
&&& F:2141 & F:2351 & $\;\;$F:2050\tablenotemark[3] & F:2357 & F:2584 \\
&&&&&&&\\
$a_3(2070)$\tablenotemark[2] & [4] & 2070$\pm$ 20 & 1926 & 1986 & 2050 & 1951 & 2002 \\
&&&&&&&\\
$a_3(2310)$ & [4] & 2310$\pm$ 40 & 2247 & 2351 & -- & 2401 & 2584 \\
&&&&&&&\\
$a_4(2260)$ & [4] & 2260 $\pm$ 15 & F:2341 & F:2351 & -- & F:2451 & F:2584 \\
&&& H:2315 & H:2492 & -- & H:2402 & H:2658 \\
&&&&&&&\\
\end{tabular}
\end{center}
\tablenotemark[1]{As the $a_0$ ground state appears at 1321 MeV in model ${\cal A}$, we identify this state with the observed $a_0(1450)$ \cite{pap47} and not with the low-lying $a_0(980)$. Therefore we assign the observed $a_0(2025)$ \cite{pap01} to the first radial excitation of the $a_0$ in model ${\cal A}$, whereas in model ${\cal B}$ the stated masses (FSM: 2071 MeV, NRM: 2250 MeV) correspond to the second radial excitation.}\\
\tablenotemark[2]{Although the $a_3(2070)$ seems to be the ground state in the 3$^{++}$-sector it does not appear in the listings of the {\sc Particle Data Group}. Its mass is much more questionable than the well established ground state masses in all other sectors up to $J = 6$. Therefore this state was excluded from the fit of the model parameters to the experimental ground state masses.}\\
\tablenotemark[3]{These masses belong to the same state. Clearly, its identification with the observed $a_2(2100)$ \cite{pap04} does not account for the {\it two} underlying $a_2$-states, $a_2(1660)$ and $a_2(1750)$, listed in the latest PDG-issue \cite{pap47}.}
\end{table}
\begin{table}
\begin{center}
\caption{New candidates for isoscalar resonances and their masses in [MeV]. The first and second columns show the name and publication reference of each resonance, respectively; the third column contains the experimental mass and its error; the calculated masses in the full Salpeter model (FSM) are shown in the fourth (model ${\cal B}$) and in the seventh (model ${\cal A}$) column. The corresponding results in their nonrelativistic reduction model (NRM) are also shown. The sixth column shows some calculations done by Godfrey and Isgur in their relativized quark model [15]. The states labeled with $\bar n n$ contain nonstrange quarks only; the states labeled with $\bar s s$ contain strange quarks only. The dominantly F-wave and H-wave states of the $f_4$ are labeled by F and H, respectively.}
\label{tab2}
\begin{tabular}{cccccccc}
&&&&&&&\\
&&& \multicolumn{2}{c}{{\bf Model ${\cal B}$}} && \multicolumn{2}{c}{{\bf Model ${\cal A}$}} \\
&&&&&&&\\
{\bf Name} & {\bf Ref.} & {\bf Exp. Mass} & {\bf FSM} & {\bf NRM} & {\bf Godfrey \& Isgur} & {\bf FSM} & {\bf NRM} \\
&&&&&&&\\
\cline{1-8}
&&&&&&&\\
$f_1(1700)$ & [7] & 1700 & $n\bar n$:1718 & $n\bar n$:1845 & $n\bar n$:1820 & $n\bar n$:1876 & $n\bar n$:1927 \\
&&& $s\bar s$:1958 & $s\bar s$:2021 & $s\bar s$:2030 & $s\bar s$:2129 & $s\bar s$:2012 \\
&&&&&&&\\
$f_1(2340)$ & [7] & 2340 $\pm$ 40 & $n\bar n$:2411 & $n\bar n$:2586 & -- & $n\bar
n$:2791 & $n\bar
n$:3073 \\
&&& $s\bar s$:2681 & $s\bar s$:2774 & -- & $s\bar s$:3094 & $s\bar s$:2954 \\
&&&&&&&\\
$\eta_2(2040)$ & [7] & 2040 $\pm$ 40 & $n\bar n$:1997 & $n\bar n$:2110 & -- & $n\bar n$:2156 & $n\bar n$:2266 \\
&&& $s\bar s$:2231 & $s\bar s$:2279 & -- & $s\bar s$:2424 & $s\bar s$:2291 \\
&&&&&&&\\
$\eta_2(2300)$ & [7] & 2300 $\pm$ 40 &$n\bar n$:2318 & $n\bar n$:2463 & -- & $n\bar n$:2593 & $s\bar s$:2753 \\
&&& $s\bar s$:2571 & $s\bar s$:2641 & -- & $s\bar s$:2890 & $n\bar n$:2829 \\
&&&&&&&\\
$f_3(2000)$ & [7] & 2000 $\pm$ 40 & $n\bar n$:1926 & $n\bar n$:1986 & $n\bar n$:2050 & $n\bar
n$:1951 & $n\bar
n$:2002 \\
&&& $s\bar s$:2128 & $s\bar s$:2134 & $\;$ $s\bar s$:2230\tablenotemark[1] & $s\bar s$:2193 & $s\bar s$:2074 \\
&&&&&&&\\
$f_3(2280)$ & [7] & 2280 $\pm$ 30 & $n\bar n$:2247 & $n\bar n$:2351 & -- & $n\bar
n$:2402 & $n\bar
n$:2584 \\
&&& $s\bar s$:2476 & $s\bar s$:2514 & $\;$ $s\bar s$:2230\tablenotemark[1] & $s\bar s$:2684 & $s\bar s$:2553 \\
&&&&&&&\\
$f_4(2320)$ & [7] & 2320 $\pm$ 30 & $n\bar n$;F:2342 & $n\bar n$;F:2351 & -- & $n\bar n$;F:2451 & $n\bar n$;F:2584 \\
&&& $n\bar n$;H:2315 & $n\bar n$;H:2492 & -- & $n\bar n$;H:2402 & $s\bar s$;F:2553 \\
&&&&&&&\\
\end{tabular}
\end{center}
\tablenotemark[1]{These masses belong to the same state.}
\end{table}

\end{document}